\documentclass[twocolumn]{aastex631} 

\usepackage{graphicx}
\usepackage{amsmath}
\usepackage{float}
\usepackage{natbib}
\usepackage{tabularx}
\usepackage{bm}
\usepackage{appendix}
\usepackage{longtable}
\usepackage{enumitem}
\usepackage{booktabs}
\usepackage{multirow}

\shorttitle{Gaia NS Mass Distribution}
\shortauthors{}

\graphicspath{{./}{figures/}}

\begin{document}

\title{Surprisingly Similar: The Mass Function of \textit{Gaia} Neutron Stars and First-Born Double Neutron Stars}

\correspondingauthor{Aryanna Schiebelbein-Zwack}
\email{aryanna.schiebelbein@mail.utoronto.ca}

\author[0009-0003-3908-6112]{Aryanna Schiebelbein-Zwack}
\affiliation{David A. Dunlap Department of Astronomy and Astrophysics, University of Toronto, 50 St George St, Toronto ON M5S 3H4, Canada}
\affiliation{Canadian Institute for Theoretical Astrophysics, 60 St George St, University of Toronto, Toronto, ON M5S 3H8, Canada}

\author[0000-0001-5484-4987]{L.~A.~C.~van~Son}
\affiliation{Center for Computational Astrophysics, Flatiron Institute, 162 Fifth Avenue, New York, NY 10010, USA}
\affiliation{Department of Astrophysical Sciences, Princeton University, 4 Ivy Lane, Princeton, NJ 08544, USA}
\affiliation{Department of Astrophysics/IMAPP, Radboud University, P.O. Box 9010, NL-6500 GL Nijmegen, The Netherlands}  

\author[0000-0002-1980-5293]{Maya Fishbach}
\affiliation{Canadian Institute for Theoretical Astrophysics, 60 St George St, University of Toronto, Toronto, ON M5S 3H8, Canada}
\affiliation{David A. Dunlap Department of Astronomy and Astrophysics, University of Toronto, 50 St George St, Toronto ON M5S 3H4, Canada}
\affiliation{Department of Physics, 60 St George St, University of Toronto, Toronto, ON M5S 3H8, Canada}

\author[0000-0003-1540-8562]{Will M. Farr}
\affiliation{Department of Physics and Astronomy, Stony Brook University, Stony Brook NY 11794, USA}
\affiliation{Center for Computational Astrophysics, Flatiron Institute, New York NY 10010, USA}

\begin{abstract}

The mass distribution of neutron stars encodes information about their formation and binary evolution. We compare the masses of two distinct populations: I) the recently identified \textit{Gaia} neutron stars in wide orbits with solar-like companions and, II) the assumed first-born recycled pulsar in Galactic double neutron star systems. 
Naively, one would expect their masses to differ due to both the presumed differences in their evolutionary histories, as well as astrophysical selection effects that can filter out configurations that would merge or be disrupted.
Yet, we find that their mass distributions are strikingly similar. Using a two-component Gaussian model, we find that both populations exhibit a narrow component centred near $1.3 \text{ M}_\odot$, accompanied by a broader, higher-mass component that extends the distribution toward larger masses. The highest density regions of their fitted parameter posteriors coincide by over 91.6\%. Statistical tests further confirm the agreement between these distributions with a Jensen-Shannon divergence $JS < 0.08$ and an earth mover's distance of $W <0.063 \text{ M}_\odot$ at 90$\%$ credibility.
This finding seems to imply that both mass functions reflect the natal mass distribution of first-born neutron stars in binary systems, supporting the hypothesis that neutron stars can be born with high masses. Consequently and perhaps surprisingly, binary evolutionary processes need not impart features on the NS mass distribution.

\end{abstract}

\section{Introduction} \label{sec:intro}
The shape of the observed mass distribution of neutron stars (NSs) is a consequence of multiple complex and still poorly understood physical processes, including stellar evolution, supernova physics, and the NS equation of state \citep{1996ApJ...457..834T, 2004ApJ...612.1044P, 2011MNRAS.414.1427V, 2012ApJ...757...55O, 2013arXiv1309.6635K, 2016arXiv160501665A, 2016ARA&A..54..401O}. The global mass distribution of known NSs is commonly modelled as a bimodal function, with a sharp peak at $\sim 1.3 \text{ M}_\odot$ and a second wider peak near $\sim 1.8 \text{ M}_\odot$ \citep{2011MNRAS.414.1427V, 2013arXiv1309.6635K,2016arXiv160501665A, 2018MNRAS.478.1377A}. More recently, a study by \citet{2025NatAs...9..552Y} found that the masses of NSs can be described by a unimodal distribution when accounting for the mass accreted by recycled pulsars in binary systems. Regardless of the true structure of the NS mass distribution, it remains unclear whether the apparent structure and the overall shape of the mass distribution reflects the natal mass function of NSs or is instead shaped by binary evolution and observational selection effects. 

NSs are found in a wide range of binary configurations, including young pulsars with OB companions \citep{1996Natur.381..584K, 2011MNRAS.412L..63B, 2014MNRAS.437.3255S, 2015MNRAS.451..581L, 2023ApJ...943...57A, 2024A&A...682A.178V}, wind-fed high-mass X-ray binaries \citep[e.g.,][]{2011Ap&SS.332....1R}, low-mass X-ray binaries with Roche-lobe–overflowing companions \citep[e.g.,][]{2024A&A...684A.124F}, recycled pulsars with white dwarf or NS companions \citep[e.g.,][]{2012MNRAS.425.1601T, 2017ApJ...846..170T}, non-interacting NS–main-sequence binaries detected astrometrically (e.g., Gaia NSs; \citealt{2024OJAp....7E..58E}), and gravitational-wave (GW) sources \citep{2017PhRvL.119p1101A, 2020ApJ...892L...3A,2021ApJ...915L...5A, 2025arXiv250818083T}. These different subpopulations of NSs contribute to the global NS mass distribution in unique ways, representing snapshots of distinct binary evolutionary stages and offering insight into their histories and interaction pathways. 

Several studies have argued that binary interactions can significantly reshape a star’s core structure, altering both the SN outcome and the properties of the resulting remnant \citep{2010ApJ...719..722S, 2021A&A...645A...5S, 2021A&A...656A..58L}. Consider electron-capture supernovae \citep[ECSNe;][]{1980PASJ...32..303M, 1984ApJ...277..791N, 1987ApJ...322..206N}, which may be more common in stars that have undergone binary stripping \citep{2004ApJ...612.1044P}.
ECSNe are expected to produce NSs with a narrower mass range and smaller natal kicks compared to iron-core collapse supernovae, which yield a broader mass distribution extending to higher masses \citep{1996ApJ...457..834T, 2004ApJ...612.1044P, 2010ApJ...719..722S, 2011MNRAS.414.1427V, 2012ApJ...757...55O, 2018MNRAS.478.1377A}.
Similarly, ultra-stripped supernovae (USSNe), following case BB mass transfer \citep{1981A&A....96..142D, 2015MNRAS.451.2123T}, are a subclass of stripped-envelope SNe that are thought to produce lower mass remnants with weak natal kicks \citep{2015MNRAS.454.3073S, 2017MNRAS.466.2085M, 2018MNRAS.479.3675M}.
Such differences in SN mechanisms and remnant properties influence the properties of surviving binaries. For instance, the gravitational wave population of NSs is proposed to originate from ECSN or USSNe due to these theoretical lower natal kicks \citep[e.g.,][]{2019MNRAS.482.2234G}. 
However, the broad, nearly uniform mass distribution (with perhaps a bias toward higher masses) seen in the gravitational wave NS population \citep{2021ApJ...921L..25L}, seems to be in contradiction with the sharp 1.25 $\text{ M}_\odot$ feature associated with ECSN \citep{2010ApJ...719..722S}.

Lastly, the origin of the more massive end of the NS mass distribution has long been a subject of particular interest. Post-formation accretion is one possible contribution; millisecond pulsars can gain $\sim 0.1-0.3 \text{ M}_\odot$ through accretion, whereas the recycled, first-borns in Galactic double NS (DNS) systems are thought to gain only $<0.02 \text{ M}_\odot$ \citep{2013arXiv1309.6635K, 2014A&A...561A..11V,2016arXiv160501665A, 2016ARA&A..54..401O, 2017ApJ...846..170T, 2021ApJ...922..158L}. Yet, accretion alone can not explain the high-mass features of the mass distribution since the difference in the $1.3 \text{ M}_\odot$ and $1.8 \text{ M}_\odot$ features exceeds the typical accreted mass required to spin-up a pulsar \citep{2013arXiv1309.6635K, 2021ApJ...922..158L}. The connection between accretion, spin-up, and mass is further complicated by observations such as those of PSR J1802-2124 and PSR J1807-2500B, which seemingly have histories of accretion due to their orbital properties and spins and yet have masses $1.24 \pm 0.11 \text{ M}_\odot$ and $1.3655 \pm 0.0021 \text{ M}_\odot$ respectively \citep{2016ARA&A..54..401O}. Hence, it is possible to spin-up a NS without making it overly massive. Growing evidence suggests that some NSs can instead be born massive depending on their final iron core mass \citep{1996ApJ...457..834T, 2011MNRAS.414.1427V, 2013arXiv1309.6635K, 2016arXiv160501665A, 2016MNRAS.460..742M, 2018ApJ...860...93S}. Still, theoretical predictions for the final remnant mass after a core collapse supernova often underpredict the location of the $1.3 \text{ M}_\odot$ peak while overpredicting its width, which is especially narrow in the case of DNSs considering stochastic processes such as fallback \citep{2012ApJ...757...55O,2016MNRAS.460..742M}, although \cite{2018ApJ...860...93S} do find their model predicts an average NS mass of $1.38 \text{ M}_\odot$. 

In short, the relationship between features in the NS mass distribution and the physical processes that shape them remains unclear, and several of the currently proposed explanations appear to contain contradictions.

In this work, we will compare two distinct NS populations — the galactic DNSs and the newly discovered NSs in \textit{Gaia} data release 3 \citep{2024OJAp....7E..58E} — to assess how their presumed distinct evolutionary histories manifest in their respective mass distributions. As implied by the previous discussion, it is well established that NSs in different systems are expected to produce different mass distributions \citep{2004ApJ...612.1044P, 2013arXiv1309.6635K}. According to isolated binary evolution models, the recycled NS in DNS binaries are the first born, analogous to the \textit{Gaia} NSs that are the only NS in their binary systems \citep{2017ApJ...846..170T}. Hence, we compare these two populations in particular because both populations are expected to represent the first-born NS in their respective binaries, and both have minimal observational biases in their mass measurements, enabling a direct comparison (see Section \ref{sec:methods}). Naively, we expect them to be very different in their evolutionary history, and thus their masses. By using the remnants of stars to piece together their evolutionary history, analogous to paleontologists who use fossils to construct phylogenetic trees, we can gain insight into the ``stellar phylogeny".

Galactic DNSs are survivors of complex evolutionary processes, including mass transfer and two supernovae. Their masses preserve this rich history, which is shaped by poorly understood physics. After the binary survives the first supernova and formation of the first-born NS, the companion star loses mass to the first-born in a mass transfer event that can shrink their orbits down to the scales of $\leq$ days \citep{2016ARA&A..54..401O, 2017ApJ...846..170T}. The first-born is referred to as `recycled' due to the spin-up that occurs during accretion, although as mentioned, these are mildly recycled compared to millisecond pulsars \citep{2014A&A...561A..11V,2016arXiv160501665A, 2016ARA&A..54..401O, 2017ApJ...846..170T}. In contrast, its companion, the second-born NS, is termed `slow'. These two types of NSs, paired in a binary, have distinct mass distributions: the recycled NSs have previously been modelled with a double Gaussian, whereas the slow NSs comparatively favour a unimodal distribution without support at masses beyond $\sim 1.55 \text{ M}_\odot$ \citep{2019ApJ...876...18F}. 

Conversely, the \textit{Gaia} population of NSs are in wide orbits with solar-like main sequence companions and therefore are not expected to have undergone a recycling process \citep{2024OJAp....7E..58E}. 
Prior to the birth of the first NS, \textit{Gaia} progenitor binaries have more extreme mass ratios and (most likely) wider orbital periods than the DNS progenitor binaries, likely leading to different mass transfer histories \citep{2024ARA&A..62...21M}. As a consequence, these two populations of NSs will have experienced different degrees of stripping. We would expect these distinctions to impact the masses of the first-born NSs and the supernova kicks imparted on their systems, thereby impacting their survivability. As well, these binaries have different observational selection effects impacting their detection (but not necessarily their masses).

In Section \ref{sec:methods}, we discuss the data that we utilise, as well as our methods for fitting a parametric form for both the \textit{Gaia} and DNSs mass distributions independently. We detail our results and investigate their statistical significance in Section \ref{sec:results}. In Section \ref{sec:discussion}, we suggest physical interpretations and implications of our results before concluding and providing a future outlook in Section \ref{sec:conclusion}.

\section{Methods} \label{sec:methods}
\subsection{Data}
The complete table of data used in this paper and its references can be found in Table \ref{tab:ns_masses}. This data is also available in the GitHub repository: \url{https://github.com/arysch/NS_mass_distributions}.

For the \textit{Gaia} NSs, we use the full sample of 21 objects identified by \citeauthor{2024OJAp....7E..58E} (\citeyear{2024OJAp....7E..58E}) using \textit{Gaia}'s third data release astrometry. This sample is restricted to objects with masses $>1.25 \text{ M}_\odot$ to prevent contamination from white dwarfs, although some contamination could still be present in the dataset \citep{2024OJAp....7E..58E}. Nonetheless, we utilise the entire sample and discuss the implications of this choice in Section \ref{sec:discussion}.

The DNSs were discovered with radio pulsar timing observations. To date, there are 26 DNSs (or candidates) contained in the Australia Telescope National Facility (ATNF) Pulsar Catalog \citep{2005AJ....129.1993M}. In addition, there are a handful of recent discoveries that have not yet been added to the ATNF catalog, such as those made by the Five-hundred-meter Aperture Spherical Telescope (FAST) \citep{2024MNRAS.530.1506S, 2025RAA....25a4003W}. In this work however, we only consider the 13 systems that have precise individual mass measurements.

Since we seek to use the mass distribution to study the impacts of evolutionary history, we focus on observed masses and do not correct for possible accretion. Therefore, we expect our inferred mass distributions to represent natal NS masses alongside any mass accretion, which we expect to be minimal for the two populations in this study.
We also neglect the impact of selection effects, similar to \citeauthor{2019ApJ...876...18F} (\citeyear{2019ApJ...876...18F}). 
Although both populations in this study are subject to their own selection effects, these are not expected to depend directly on mass, aside from the $>1.25 \text{ M}_\odot$ cut applied to the \textit{Gaia} sample. The dominant selection effect for the \textit{Gaia} NS sample is the orbital period — binaries with $P\sim 1-10^3$ days are most identifiable with \textit{Gaia} \citep{2024NewAR..9801694E}. 
In DNS systems, selection biases make it particularly challenging to detect very tight and short-period binaries \citep{2016ARA&A..54..401O, 2011MNRAS.413..461O}. These selection effects only depend on mass indirectly (e.g. through possible evolutionary correlations between mass and period) and thus we fit a distribution to the observed mass measurements without any correction \citep{2004hpa..book.....L, 2013arXiv1309.6635K}. 

\subsection{Fitting mass distributions}

Many previous works have focused on fitting a functional form for the mass distribution of NS subpopulations \citep[e.g,][]{2011MNRAS.414.1427V, 2012ApJ...757...55O,2013arXiv1309.6635K,2016arXiv160501665A,  2018MNRAS.478.1377A, 2019ApJ...876...18F,2019MNRAS.485.1665K, 2020RNAAS...4...65F, 2025NatAs...9..552Y}. A double Gaussian model has commonly been used to capture the peak at low-masses ($\sim 1.3 \text{ M}_\odot$) and the high-mass ($\gtrsim 1.4 \text{ M}_\odot $) tail for NS systems \citep{2010ApJ...719..722S,2013arXiv1309.6635K, 2016arXiv160501665A, 2018MNRAS.478.1377A, 2019ApJ...876...18F, 2020RNAAS...4...65F}. While alternative functional forms for the NS birth-mass distribution have been proposed (e.g. \citealt{2025NatAs...9..552Y}, which did not include the \textit{Gaia} systems), the present analysis is not intended to adjudicate between competing models of the underlying distribution. Instead, the focus is on comparing the mass distributions inferred for the two populations, based on models previously analysed in the literature.

To determine the parametric mass distribution of the \textit{Gaia} NS population we use the two-component model

\begin{equation}\label{eq:model}
\begin{split}
    p(m|\Lambda) &= p(m|\{\mu_1, \mu_2, \sigma_1, \sigma_2, \omega\}) \\
    & = \omega \cdot \mathcal{N}(m \mid \mu_1, \sqrt{\sigma_1^2 + \sigma_{obs}^2}) +\\
    & (1 - \omega) \cdot \mathcal{N}(m \mid \mu_2, \sqrt{\sigma_2^2 + \sigma_{obs}^2})\\
\end{split}
\end{equation}

where $\mathcal{N}$ is the Gaussian or normal distribution, $\mu_i$ are the means of the Gaussians, $\sigma_i$ are their widths, $\omega$ is the relative weight of the two Gaussians, and $\sigma_{obs}$ are the measurement uncertainties on the observational data. 

Our prior assumptions (given in Table \ref{t:priortable}, also with the results) are uniform distributions between $0 \text{ M}_\odot$ and $2.1 \text{ M}_\odot$ for both the means and standard deviations of the Gaussians; $\mu_1$, $\mu_2$, $\sigma_1$, $\sigma_2$. The mean of the second Gaussian, $\mu_2$, is additionally constrained to be a larger value than the first, $\mu_2>\mu_1$. The weight parameter, $\omega$, prior is uniformly distributed between 0 and 1.

\begin{table}[ht!]
\centering 
\caption{The model parameters, their prior assumptions, and the resulting fit to both \textit{Gaia} NS's and the recycled NS in galactic DNS systems. The median and $90\%$ credible uncertainties are reported. Both populations parameters are in agreement.}
\label{t:priortable}
\begin{tabular}{cccc}
\toprule
Parameter & Prior & \textit{Gaia} NS & Recycled DNS \\
\midrule\midrule
  $\mu_1$           &$\mathrm{U}(0.0, 2.1) $    & $1.36^{+0.04}_{-0.04}$ & $1.34^{+0.07}_{-0.05}$ \\
  $\sigma_1$        & $\mathrm{U}(0.0, 2.1)$    &  $0.05^{+0.07}_{-0.04}$ &$0.02^{+0.22}_{-0.02}$ \\
  $\mu_2$           & $\mathrm{U}$($\mu_1$, 2.1)     & $1.65^{+0.19}_{-0.29}$ & $1.51^{+0.19}_{-0.18}$ \\
  $\sigma_2$        & $\mathrm{U}(0.0, 2.1)$    & $0.21^{+0.24}_{-0.18}$ & $0.14^{+0.33}_{-0.14}$ \\
  $w$               & $\mathrm{U}(0, 1)$         & $0.74^{+0.23}_{-0.27}$ & $0.68^{+0.32}_{-0.35}$\\
\midrule\midrule
\end{tabular}
\end{table}

We use \texttt{NumPyro} and \texttt{JAX} with the NUTS sampler to infer these parameters \citep{2011arXiv1111.4246H,jax2018github, 2018arXiv181009538B,2019arXiv191211554P}. The code used for this project is available in the GitHub repository: \url{https://github.com/arysch/NS_mass_distributions}.

\subsection{Statistical Comparison}

To quantitatively compare the mass distributions of the \textit{Gaia} and recycled DNS, we utilise the Jensen-Shannon (JS) divergence and the Wasserstein metric.

The JS divergence is a symmetrised and bounded version of the Kullback–Leibler divergence (KL) defined by

\begin{equation}\label{eq:JSD}
\begin{split}
    \text{JS}[p(x), q(x)] &= \frac{\text{KL} [p(x), \frac{p(x)+q(x)}{2}] +\text{KL} [q(x), \frac{p(x)+q(x)}{2}]}{2},
\end{split}
\end{equation}
where $p,q$ are the distributions over $x$ that are compared \citep{10.1214/aoms/1177729694, 61115}. The values of $JS$ can vary between 0 and 1, where 0 indicates the distributions are identical and 1 indicates the distributions are maximally divergent. In units of bits, this describes how much information is needed on average to describe one population using the other.

The Wasserstein metric, or earth mover's distance, measures how different two mass distributions are in terms of the amount of mass that would need to be shifted from one to make them match \citep{Villani2009}. It is defined by

\begin{equation}\label{eq:W}
    W[p(x),q(x)] = \int^{\infty}_{-\infty} |P(x) - Q(x)|dx
\end{equation}

where $P,Q$ are the cumulative distributions of the distributions $p,q$ defined over $x$.

\section{Results} \label{sec:results}
\subsection{Comparing Our Inference to Previous Results}
\begin{figure*}
    \centering
    \includegraphics[width=\linewidth]{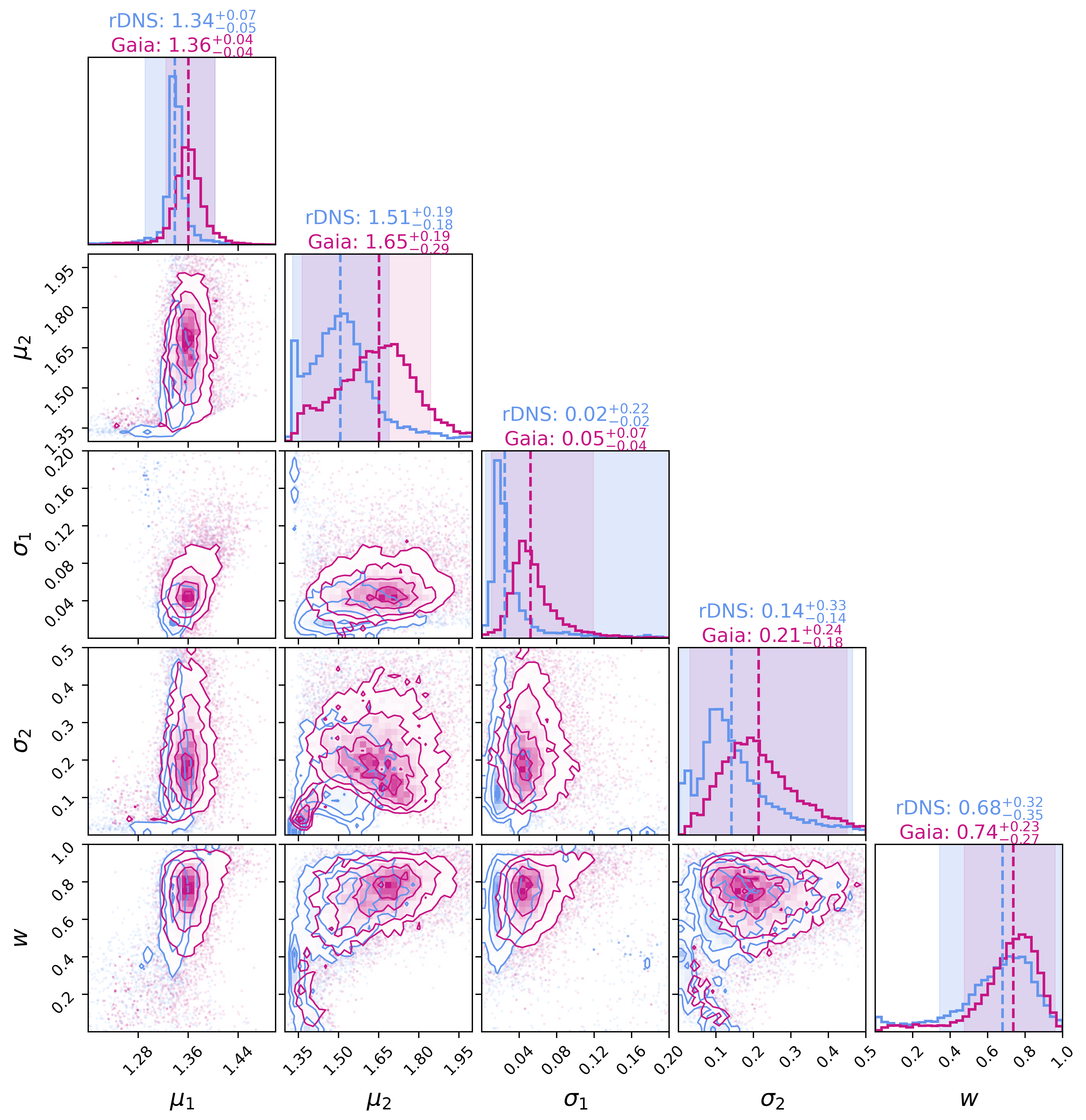}
    \caption{A corner plot showing the posteriors of 10,000 samples of the double Gaussian model for each of the \textit{Gaia} NS population (pink) and the recycled DNS population (blue). The numeric values shown are the median and $90\%$ credible limits (which are also provided in Table \ref{t:priortable}). The dotted lines indicate the respective median values and the banded regions show the respective HDIs. The two populations' posteriors overlap indicating their distributions are similar. The HDI for the mean of the second Gaussian, $\mu_2$, extends to larger masses for the \textit{Gaia} population. The weight parameter, $\omega$, has low support at values of $\omega = 0,1$, indicating preference for a double Gaussian model.}
    \label{fig:corner}
\end{figure*}

Our recycled DNS mass distribution fits are in successful agreement with those of \citeauthor{2019ApJ...876...18F} (\citeyear{2019ApJ...876...18F}) despite using a smaller sample size with only precise individual mass measurements. Namely, we find evidence for a peak at $\mu_1 =1.34^{+0.07}_{-0.05} \text{ M}_\odot$ with a standard deviation of $\sigma_1 = 0.02^{+0.22}_{-0.02}$ while they find $\mu_1 =1.34^{+0.01}_{-0.01} \text{ M}_\odot$ and $\sigma_1 =0.02^{+0.04}_{-0.01} \text{ M}_\odot$. Additionally, we find support for a second Gaussian to be necessary to describe the high mass data with a weight parameter $w=0.68^{+0.32}_{-0.35}$ compared to their $w=0.68^{+0.16}_{-0.22}$. Our mass distribution model is nested: when 
$w=0,1$, the two-component Gaussian mixture collapses to a single Gaussian. Posterior support away from these boundary values indicates evidence for the bimodal model, as seen in the 1D marginal plot of $w$ in Figure \ref{fig:corner}. This second Gaussian has parameters $\mu_2=1.51^{+0.19}_{-0.18} \text{ M}_\odot$ and $\sigma_2 = 0.14^{+0.33}_{-0.14} \text{ M}_\odot$ compared to the \citeauthor{2019ApJ...876...18F} (\citeyear{2019ApJ...876...18F}) values of $\mu_2=1.47^{+0.11}_{-0.09} \text{ M}_\odot$ and $\sigma_2 = 0.15^{+0.11}_{-0.06} \text{ M}_\odot$. The results of our parameter estimation are summarized in Table \ref{t:priortable}, as well as the corner plot in Figure \ref{fig:corner}. 

We emphasise that the necessity of the second Gaussian does not equate to the need for a second 'peak'. As showcased in Figure \ref{fig:mass_dists}, the mass distributions have an extended tail beyond $1.5 \text{ M}_\odot$ but not necessarily a prominent peak. As mentioned previously, model specification is beyond the scope of this work. The possibility of a unimodal model is not ruled out but strongly disfavored, as can be seen from the small but still present draws near $w=0,1$ which correspond to the $\mu_2$ draws that are close to the median $\mu_1 \simeq 1.3 \text{ M}_\odot$, indicating the superposition of both Gaussians. Table \ref{t:hdi} in the Appendix contains the highest density interval (HDI) that covers 90\% of the posterior for each parameter, and also shows that $w=1$ is contained in the HDI for the recycled DNS sample. However, this is likely due to the smaller sample size compared to \textit{Gaia} and an analysis of both populations combined does constrain $w<0.88$. 

\subsection{Comparing the \textit{Gaia} and Recycled DNS Populations}

\begin{figure*}[t]
    \includegraphics[width=0.49\textwidth]{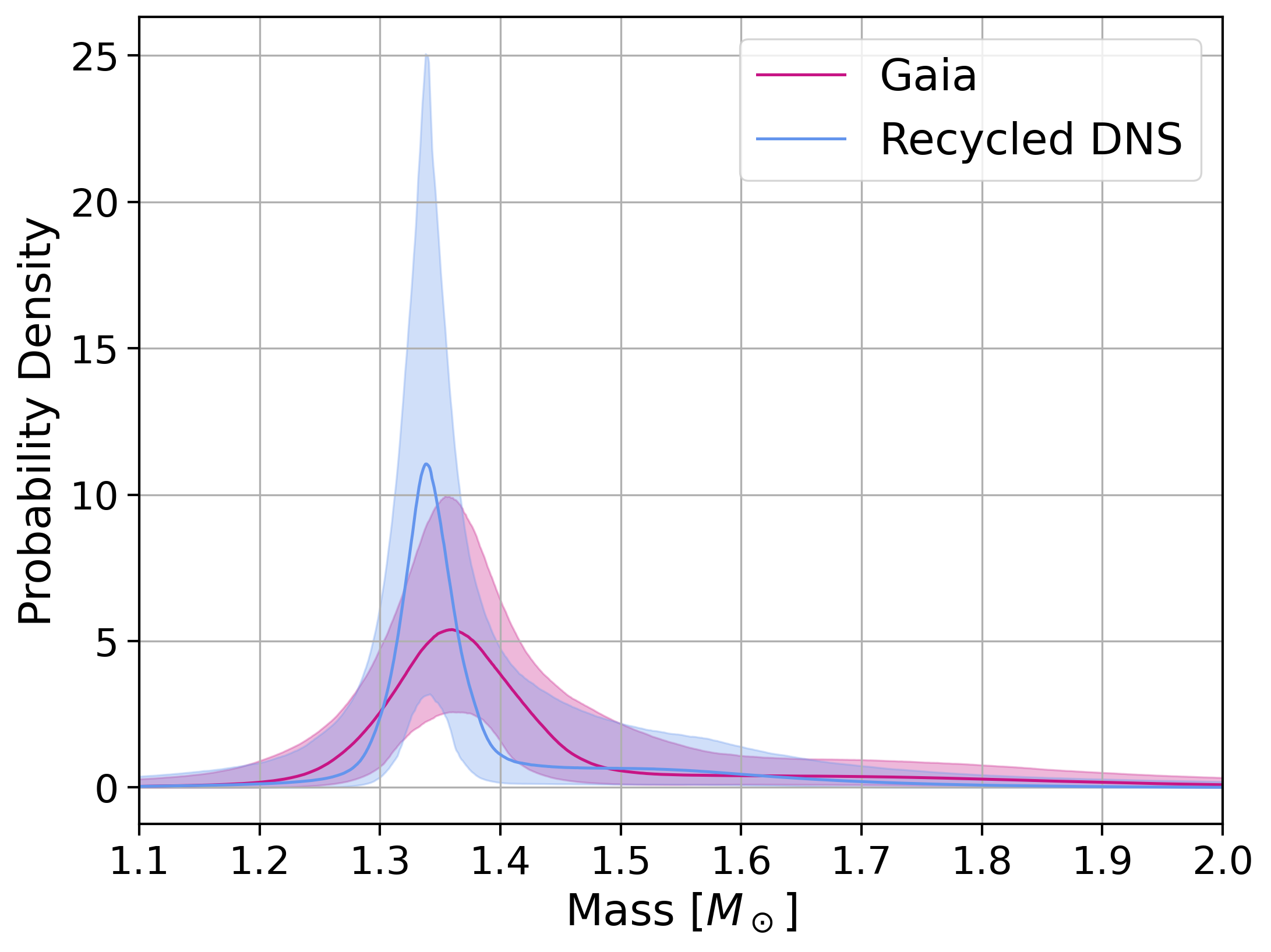}
    \includegraphics[width=0.49\textwidth]{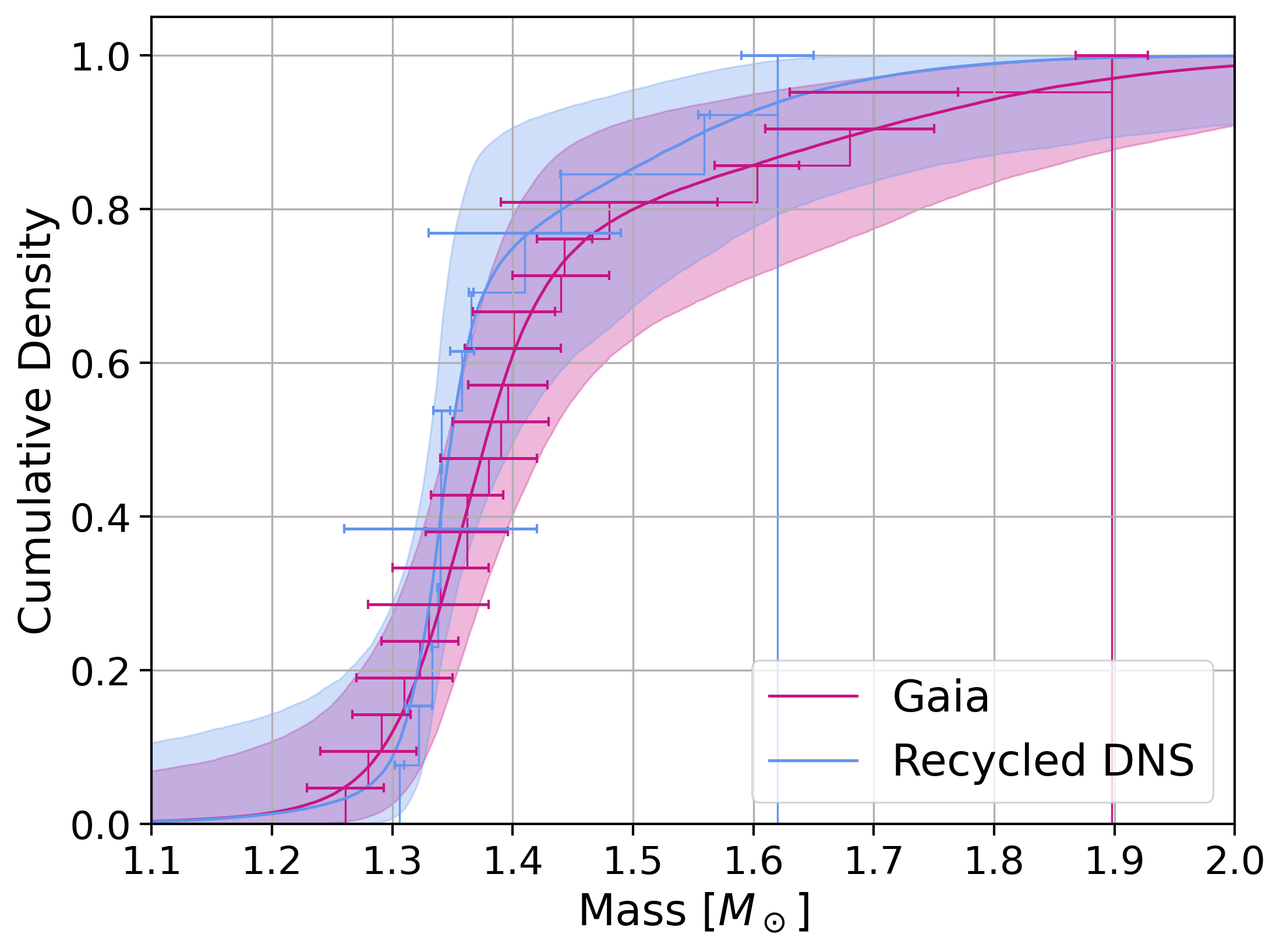}

    \caption{\textit{Left:} Posterior predictive distributions of the double Gaussian model fit to the \textit{Gaia} NSs (pink) and the recycled DNSs (blue). The bands shown depict the $90\%$ credible regions. The second Gaussian is necessary compared to a single Gaussian due to the tail towards high masses, but is not to be mistaken for a peak. \textit{Right:} Cumulative posterior predictive distributions shown with data points overplotted.}
    \label{fig:mass_dists}
\end{figure*}

The \textit{Gaia} NS population parameters are remarkably consistent with those of the recycled DNSs. This can be seen visually in Figures \ref{fig:corner} and \ref{fig:mass_dists}. In Figure \ref{fig:corner}, it is clear that there is significant overlap in the model parameters of each population. In Figure \ref{fig:corner} (see Table \ref{t:hdi} for numeric values) the HDIs reported for each model parameter have remarkable overlap computed in relation to the range of the tightest posterior;  $\mu_1, \sigma_1, \sigma_2,$ and $w$ all have 100\% overlap with the HDI of the \textit{Gaia} population completely enveloped by that of the first-born DNSs. The exception, $\mu_2$, has 91.6\% overlap with the \textit{Gaia} HDI extending to higher masses than the recycled DNSs.

Notably, while the \textit{Gaia} HDIs overlap almost completely with those of the DNSs, they are also much narrower (owing to the larger sample size of the \textit{Gaia} data) and their medians skew larger than those of the DNS population. We find that for all model parameters, although their HDIs overlap more than $90\%$, the \textit{Gaia} posteriors are larger than the DNS posteriors with the following credibilities:  

$$P(\mu_1^{\textit{Gaia}}>\mu_1^{\text{DNS}})=80\%$$ 
$$P(\mu_2^{\textit{Gaia}}>\mu_2^{\text{DNS}})=75\%$$
$$P(\sigma_1^{\textit{Gaia}}>\sigma_1^{\text{DNS}})=75\% $$
$$P(\sigma_2^{\textit{Gaia}}>\sigma_2^{\text{DNS}})=66\%$$ 
$$P(w^{\textit{Gaia}}>w^{\text{DNS}})=60\%$$

In Section \ref{sec:combined}, we present a joint inference of the total mass distribution from the combined data sets and examine the resulting HDIs, which show higher constraining power due to the increased effective sample size.

To further test whether the \textit{Gaia} NS mass distribution extends to higher masses than the recycled DNS population, we compare posterior cumulative distribution function (CDF) support (Figure~\ref{fig:mass_dists}). At masses $>1.4 \text{ M}_\odot$ ($>1.6 \text{ M}_\odot$), the \textit{Gaia} posterior assigns between $21\text{--}60\%$ ($5\text{--}29\%$) of the total probability mass, whereas the primary DNSs assign $10\text{--}50\%$ ($1\text{--}22\%$). The percentage of support is larger in the \textit{Gaia} population at 78.5(76.3)\% credibility.

\subsubsection{Statistical Comparison}

Figures \ref{fig:corner} and \ref{fig:mass_dists} both display the similarity of the \textit{Gaia} NS and recycled DNS masses visually. Quantitatively, we assess the differences between these mass distributions with the JS divergence, described in Section \ref{sec:methods}.

We calculate $$\text{JS}[\bar p_\text{\textit{Gaia}}(M_\text{NS}),\bar p_\text{recycledDNS}(M_\text{NS})] \leq 0.08$$ at the 90\% credible level, where $\bar p$ indicates the mean of the distribution samples. This value is much closer to 0 than to 1; thus we can conclude that the distributions are similar.

Moreover, we determine that the Wasserstein distance between the \textit{Gaia} and the recycled DNS mass distributions is 
$$ W[ p_\text{\textit{Gaia}}(M_\text{NS}), p_\text{recycledDNS}(M_\text{NS})] \leq 0.063 \text{ M}_\odot$$
at $90\%$ credibility. In words, at most $0.063 \text{ M}_\odot$ would need to be redistributed to make these distributions equal. This highlights the similarity of the distributions; although the mass distributions in Figure \ref{fig:mass_dists} are defined beyond the range shown, we can restrict our range of interest between $1.2-2.0 \text{ M}_\odot$ for which the largest possible distance is $W=0.8\text{ M}_\odot$. Our computed value is nearly an order of magnitude smaller than this conservative maximum.

\section{Discussion}\label{sec:discussion}

Despite having different binary properties, formation histories, and observational selection effects, the mass distributions of recycled DNSs and \textit{Gaia} NSs are statistically indistinguishable, and are in fact surprisingly similar to each other. If this consistency persists with additional data, this could indicate there is a universality to the masses of first-born NSs. This result strengthens the case for the hypothesis that the mass distribution structure $\gtrsim 1.5 \text{ M}_\odot$ is not a consequence of similar evolutionary history, such as accretion, but could be a reflection of birth masses \citep{2013arXiv1309.6635K, 2016arXiv160501665A, 2017ApJ...846..170T}. This is additionally remarkable due to the strong resemblance between the two analysed distributions, which further supports the aforementioned insignificant accretion of the recycled DNSs. The concordance between the two mass distributions can perhaps be explained by a shared physical characteristic of the recycled DNSs and the \textit{Gaia} NSs; their shared history as the first NS born in their respective systems. 

As is the case for all NS populations, the presence and prominence of the $\mu_1 \sim 1.3 \text{ M}_\odot$ bump is explained by the typical expected NS birth mass. This typical mass is roughly set by the Chandrasekhar limit, the mass scale at which electron degeneracy can no longer support against core collapse, which depends on the composition and structure of the core \citep{1996ApJ...457..834T, 2013arXiv1309.6635K, 2016arXiv160501665A}, with additional dependence on explosion and fallback physics \citep{2002RvMP...74.1015W}. 

The mass distribution structure beyond $>1.5 \text{ M}_\odot$ could also be an artifact of the natal birth masses of NSs, since neither of the considered systems are expected to have accreted much mass, as specified in Section \ref{sec:intro}. The degree of hydrogen envelope stripping that may have occurred for each of these populations could influence their final core and remnant masses. The birth mass of a NS is largely determined by the mass of the progenitor's final iron core and the compactness, which are partially influenced by the carbon abundance earlier in stellar evolution \citep{1996ApJ...457..834T, 2004ApJ...612.1044P, 2012MNRAS.425.1601T, 2016arXiv160501665A, 2024ApJ...964L..16B}. A lower carbon abundance in the core leads to radiative carbon burning, which permits faster core contraction and ultimately the formation of a more massive iron core. In contrast, a higher carbon abundance supports convective carbon burning, which slows contraction and results in a smaller iron core. The $^{12}\text{C}/^{16}\text{O}$ ratio is established during helium burning via competition between the $\alpha-$capture reaction and the triple$-\alpha$ process, although precise predictions for this ratio are uncertain \citep[e.g.,][]{2017RvMP...89c5007D}. Retention of the hydrogen envelope allows the helium core to grow, which typically results in a lower $^{12}\text{C}/^{16}\text{O}$ ratio and thus favours conditions that produce a larger, more compact iron core \citep{2016arXiv160501665A, 2021A&A...656A..58L,2021A&A...645A...5S, 2023A&A...671A.134A}. This hypothesis can also explain one of the very slight hints of a divergence in these mass distributions; the \textit{Gaia} NS mass CDF has more support $>1.4 \text{ M}_\odot$ to 78.5\% credibility (see Section \ref{sec:results}). Progenitors of recycled DNSs in short-period binaries are more likely to have lost their envelopes due to binary interactions, limiting helium core growth and leading to smaller, less compact iron cores. In contrast, the \textit{Gaia} NS progenitors in wider systems may have retained their envelopes longer and thus produced more massive NSs at birth. This may also explain why the second-born, slow DNSs are even less massive than their recycled counterparts with masses fit to a uniform distribution $<1.55 \text{ M}_\odot$ as noted in Section \ref{sec:intro}; they experienced more stripping and thus had even less massive cores \citep{2015MNRAS.451.2123T, 2016ARA&A..54..401O}. 

There are additional, more exotic explanations for the locations and shape of features in the mass distributions. Hierarchical triples or secondary exchange encounters have been invoked to explain properties other than the NS masses \citep[i.e., the eccentric orbits of the \textit{Gaia} systems][]{2014A&A...561A..11V, 2016ARA&A..54..401O}. The \textit{Gaia} NSs are not the first case where a NS and its companion are difficult to reconcile with a primordial binary origin. \citeauthor{2014A&A...561A..11V} (\citeyear{2014A&A...561A..11V}) consider secondary exchange encounters to explain binaries in which a NS could not have been recycled by its companion, such as PSR B2127+11C or PSR B1718-19 \citep[see also ][]{1991ApJ...374L..41P,2012ApJ...745..109L}. These encounters would likely only occur in cluster environments, and 3 of the \textit{Gaia} NSs are in the galactic halo, which could indicate an origin in a disrupted cluster. Nevertheless, these explanations fall short of reconciling the notable similarity in the mass distributions between the two populations. 

Lastly, we note the caveat that not all of the \textit{Gaia} NS sample are necessarily NS, and could instead contain white dwarf contaminants \citep{2024OJAp....7E..58E}. However, we take the agreement between mass distributions to indicate this contamination is minimal if present at all. 

\section{Conclusion}\label{sec:conclusion}

We compare the mass distributions of the newly discovered population of NSs from \textit{Gaia}'s third data release with the recycled NSs in NS binaries using a double Gaussian model given by Eq. \ref{eq:model}. The \textit{Gaia} population have much wider orbits than the DNSs and also contain a less massive companion. In combination, these characteristics change the conditions for mass transfer, which are a priori expected to impact the NS masses. The mass of the NS is dependent on the final profile and structure of the iron core of its progenitor, which in turn is shaped by its ZAMS mass, metallicity, rotation, and binary properties, which need not be the same for these NSs \citep{2004ApJ...612.1044P, 2016arXiv160501665A}. 

In our comparison of the mass distributions of these two NS populations we find that:

\begin{itemize}
    \item the \textit{Gaia} NS and first-born DNS mass distributions are very similar, with population parameters in agreement as shown in Figure \ref{fig:corner}, and discussed in Section \ref{sec:results} with HDIs overlapping 100\% for nearly all model parameters ($\mu_1, \sigma_1, \sigma_2, w$; $\mu_2$ has 91.6\% overlap), $JS<0.08$, and $W<0.063 \text{ M}_\odot$

    \item neither population of NSs is expected to have accreted substantial mass due to their binary configurations, and their resemblance supports previous findings that recycled DNSs require only modest mass accretion to be spun up \citep{2013arXiv1309.6635K, 2017ApJ...846..170T, 2021ApJ...922..158L}

    \item the fraction of support in the mass distribution above $1.6 \text{ M}_\odot$ across both populations is 1–29\%, and the lack of an expected accretion history indicates NSs can be born massive, consistent with earlier results \citep[e.g.][]{1996ApJ...457..834T, 2011MNRAS.414.1427V, 2013arXiv1309.6635K, 2016arXiv160501665A}

    \item the agreement of the masses in the two independent NS datasets suggests that there are no white dwarf contaminants in the \textit{Gaia} data

    \item the nested two-component Gaussian model fit to the masses prefer a bimodal distribution compared to a unimodal distribution for both NS populations, as seen in Figure \ref{fig:corner} and discussed in Section \ref{sec:results} (see also \ref{sec:combined})

    \item we find mild evidence that the Gaia NS mass distribution assigns greater probability to masses above $1.4\text{ M}_\odot$ (Figure \ref{fig:mass_dists} and Section \ref{sec:results}), though this preference is only supported at 78.5\% credibility with current datasets
\end{itemize}

These results defy expectations in some aspects; the dissimilarities of the \textit{Gaia} and first-born DNS binary configurations are expected to be tied to dissimilarities in their masses. As highlighted repeatedly in this work, their masses could have been influenced by their discrepant mass ratios, orbital periods, kicks, and stripping or accretion histories. Yet, their mass distributions are remarkably resemblant. We propose that this indicates that the masses of first-born NSs are determined in a more universal process than previously thought. Conceivably, the distributions fit in this work (see also Section \ref{sec:combined}) could represent the natal first-born NS mass distribution, consistent with prior assumptions from the literature \citep{2012ApJ...757...55O}.

Theoretical explanations for the NS mass distributions will also leave imprints on other parameters such as periods, spins, eccentricities, natal kicks,  etc., which should be taken into account holistically in future studies. A comprehensive understanding of the stellar phylogeny also requires placing these NS binaries in the broader context of all types of stellar binary populations. Millisecond pulsars and X-ray binaries, for instance, can shed light on the impact of accretion and the high-mass end of the mass distribution. At the low-mass end of the NS mass distribution, further exploration is needed to investigate possible connections between the \textit{Gaia} population and the slow DNSs. The similarity of the \textit{Gaia} NSs to the recycled DNSs automatically precluded their similarity to the slow NSs, since previous work showed these mass distributions were different \citep{2019ApJ...876...18F}. However, the \textit{Gaia} NS sample was limited to masses $>1.25 \text{ M}_\odot$ and could potentially exclude NS that overlap with the slow NS mass range. This comparison could be insightful on low-mass NSs in wide vs. tight orbits since there is a selection effect against observing DNSs in wide orbits due to the lack of Shapiro delay \citep{2017ApJ...846..170T}. Finally, spider pulsars, which like the \textit{Gaia} NSs also have a low-mass stellar companion but are on incredibly tight orbits, provide yet another region of binary parameter space that should be explored to understand the full diversity of NS pathways \citep{2013IAUS..291..127R, 2013ApJ...775...27C}.

Additionally, while we adopt a two-component Gaussian model in this work due to common practice, there is no particular reason that Nature must adhere to this simple form. Recently, \citeauthor{2025NatAs...9..552Y} (\citeyear{2025NatAs...9..552Y}) found that their 'turn-on' power-law distribution provided a better fit to the global NS mass distribution compared to the double Gaussian. Their analysis omitted the \textit{Gaia} NSs and it remains to be seen if they are more accurately described with this model. Identifying an accurate representation of the mass distribution is essential for extracting deeper physical insight.

These proposed studies will be enriched by upcoming data of compact objects. Naturally, we can expect a larger dataset of NSs from the upcoming \textit{Gaia} data releases, DR4 in 2026 and DR5 in the 2030s, which will also be able to probe a wider parameter space \citep{2025arXiv250301533B}. There remains the possibility of probing isolated NS masses, undisturbed by binary interactions from microlensing events \citep{2024MNRAS.531.2433S}. Radio telescopes such as the Square Kilometre Array, FAST, and the Canadian Hydrogen Intensity Mapping Experiment are expected to increase the DNS sample to over 100 binaries \citep{2015aska.confE..40K,2024ApJ...966...26T, 2025RAA....25a4003W}. Gravitational wave observations have already revealed a handful of NSs, with more anticipated in upcoming observing runs of the LIGO-VIRGO-KAGRA collaboration \citep{2021ApJ...921L..25L}. Future detectors, such as LISA and Einstein Telescope, will continue to increase the number of known NSs \citep{2019MNRAS.483.2615K,2020MNRAS.492.3061L, 2025arXiv250519962S}. As these multi-messenger datasets grow, we will come closer to disentangling the entire unified phylogenetic tree of stellar evolution.

\section*{Acknowledgments}
\begin{acknowledgments}
ASZ acknowledges the support of the Natural Sciences and Engineering Research Council of Canada - Canada Graduate Scholarships - Doctoral (NSERC-CGS-D) program. 
LvS acknowledges support from the Dutch Research Council (NWO) through the NWO Talent Programme (Veni, Grant DOI: \url{https://doi.org/10.61686/XVIAV86753}).
ASZ, LvS, and MF thank the Lorentz Center and the organisers of the workshop `Challenges and future perspectives in gravitational-wave astronomy: O4 and beyond' for igniting conversations on this topic.
\end{acknowledgments}

\software{ 
NumPy \citep{harris2020array}, SciPy \citep{virtanen2020scipy}, Matplotlib \citep{hunter2007matplotlib}, pandas \citep{mckinney2010data}, Seaborn \citep{waskom2021seaborn}, JAX \citep{jax2018github}, NumPyro \citep{2019arXiv191211554P}, ArviZ \citep{kumar2019arviz}, and corner.py \citep{foreman2016corner}.}

\appendix
\renewcommand\thetable{\thesection.\arabic{table}} 
\renewcommand\thefigure{\thesection.\arabic{figure}} 
\renewcommand\theequation{\thesection.\arabic{equation}} 
\setcounter{table}{0}
\setcounter{figure}{0}
\setcounter{equation}{0}


\section{Gaia and Recycled Double Neutron Star Data}\label{appendix}
\begin{figure*}
    \centering
    \includegraphics[width=1\linewidth]{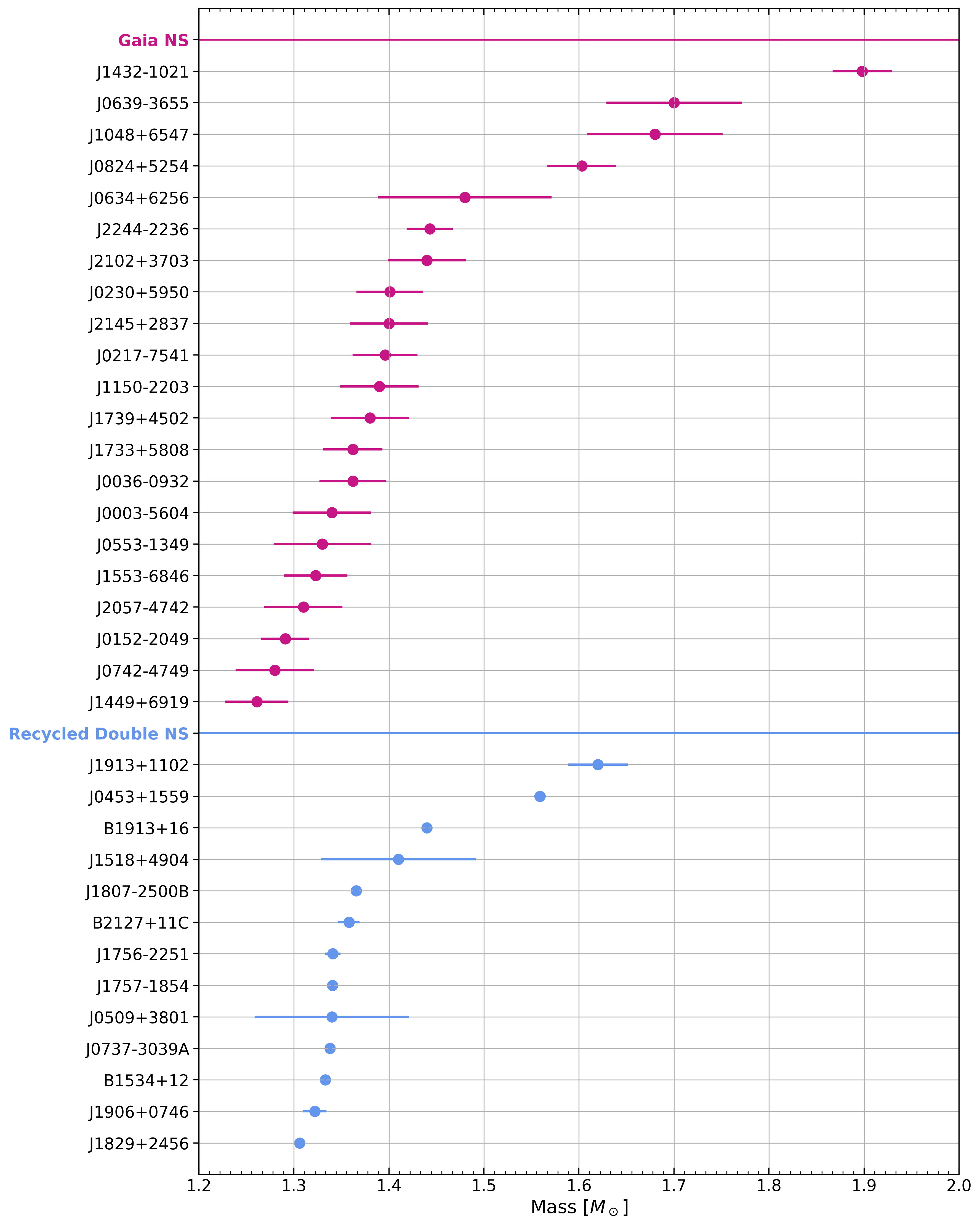}
    \caption{The masses of NSs used in this work. The \textit{Gaia} mass measurement range extends beyond the recycled DNSs, although we are subject to small-number statistics. }
    \label{fig:data}
\end{figure*}

In Figure \ref{fig:data} we present the table of data used in this work. This \textit{Gaia} data was obtained from \citeauthor{2024OJAp....7E..58E} (\citeyear{2024OJAp....7E..58E}) and the DNS data was obtained from the catalog published by \citeauthor{2019ApJ...876...18F} (\citeyear{2019ApJ...876...18F}) and supplemented with the ATNF Pulsar Catalog \citep{2005AJ....129.1993M}. We queried the ATNF catalog for pulsars with NS companions and only selected those that had a confirmed NS companion and individual mass measurements.

\begin{table*}
\centering
\caption{Mass measurements of NSs used in this work. In the \textit{Gaia} category, $m_1$ refers to the NS mass and $m_2$ refers to the stellar companion mass. All systems in this category have the same reference. In the DNS category, $m_1$ is the mass of the recycled, first-born NS and $m_2$ is the mass of the slow, second-born.}
\begin{tabular}{lccccl}
\hline
Name & $m_1$ [$\text{ M}_\odot$] & $m_2$ [$\text{ M}_\odot$] & Period [days] & Reference \\
\hline
\hline
\multicolumn{5}{l}{\textbf{\textit{Gaia} }} \\
\hline
J1432$-$1021 & $1.898 \pm 0.030$ & $0.79 \pm 0.03$ & $730.9 \pm 0.5$ & \cite{2024OJAp....7E..58E} \\
J0639$-$3655 & $1.70 \pm 0.07$ & $1.32 \pm 0.06$ & $654.6 \pm 0.6$ & -- \\
J1048+6547 & $1.68 \pm 0.07$ & $1.000 \pm 0.05$ &  $814.2 \pm 3.4$ & -- \\
J0824+5254 & $1.603 \pm 0.035$ & $1.10 \pm 0.03$ & $1027 \pm 4$ & --  \\
J0634+6256 & $1.480 \pm 0.09$ & $1.18 \pm 0.06$ & $1046 \pm 2.1$ & -- \\
J2244$-$2236 & $1.443 \pm 0.023$ & $1.002 \pm 0.030$ & $938.3 \pm 0.5$ & --  \\
J0230+5950 & $1.401 \pm 0.034$ & $1.114 \pm 0.030$ & $1029 \pm 5$ & --  \\
J2145+2837 & $1.40 \pm 0.04$ & $0.95 \pm 0.05$ & $889.8 \pm 1.2$ & --  \\
J2102+3703 & $1.44 \pm 0.04$ & $1.03 \pm 0.03$ & $480.9 \pm 0.6$ & --  \\
J0217$-$7541 & $1.396 \pm 0.033$ & $0.996 \pm 0.033$ & $636.1 \pm 0.7$ & --  \\
J1150$-$2203 & $1.39 \pm 0.04$ & $1.18 \pm 0.06$ & $631.80 \pm 0.23$ & --  \\
J1739+4502 & $1.38 \pm 0.04$ & $0.778 \pm 0.030$ & $657.4 \pm 0.6$ & --  \\
J1733+5808 & $1.362^ \pm 0.030$ & $1.16 \pm 0.05$ & $570.94 \pm 0.31$ & --  \\
J0036$-$0932 & $1.362 \pm 0.034$ & $0.94 \pm 0.04$ & $719.8\pm 0.9$ & -- \\
J0003$-$5604 & $1.34 \pm 0.04$ & $0.802 \pm 0.030$ & $561.83 \pm 0.29$ & --  \\
J0553$-$1349 & $1.33 \pm 0.05$ & $0.98  \pm 0.06$ & $189.10 \pm0.05$ & --  \\
J1553$-$6846 & $1.323 \pm 0.032$ & $1.04 \pm 0.05$ & $310.17 \pm 0.11$ & -- \\
J2057$-$4742 & $1.31 \pm 0.04$ & $1.048 \pm 0.031$ & $230.15 \pm 0.07$ & --  \\
J0742$-$4749 & $1.28 \pm 0.04$ & $0.90 \pm 0.05$ & $497.6\pm0.4$ & --  \\
J1449+6919 & $1.261 \pm 0.032$ & $0.91 \pm 0.05$ & $632.76\pm0.22$ & --  \\
\hline\hline
\multicolumn{5}{l}{\textbf{DNS}} \\
\hline
J1913+1102 & $1.620 \pm 0.03 $ & $1.270 \pm 0.03$ & 0.206 & \cite{2020Natur.583..211F} \\
J0453+1559 & $1.559 \pm 0.005$ & $1.174 \pm 0.004$ & 4.072 & \cite{2015ApJ...812..143M} \\
J1757$-$1854 & $1.3406 \pm 0.0005$ & $1.3922 \pm 0.0005$ & 0.184 & \cite{2023mgm..conf.3774C} \\
J0737$-$3039A & $1.3381 \pm 0.0007$ & $1.2489 \pm 0.0007$ & 0.102 & \cite{2006Sci...314...97K} \\
J1756$-$2251 & $1.341 \pm 0.007$ & $1.230 \pm 0.007$ & 0.320 & \cite{2014MNRAS.443.2183F} \\
B1913+16 & $1.4398 \pm 0.0002$ & $1.3886 \pm 0.0002$ & 0.323 & \cite{2010ApJ...722.1030W} \\
B1534+12 & $1.333 \pm 0.0002$ & $1.3455 \pm 0.0002$ & 0.421 & \cite{2014ApJ...787...82F} \\
B2127+11C & $1.358 \pm 0.01$ & $1.354 \pm0.01$ & 0.335 & \cite{2006ApJ...644L.113J} \\
J1518+4904 & $1.41 \pm 0.08$ & $1.31 \pm 0.08$ & 8.634 & \cite{2008AA...490..753J} \\
J0509+3801 & $1.34 \pm 0.08$ & $1.46 \pm 0.08$ & 0.380 & \cite{2018ApJ...859...93L}\\
J1807$-$2500B & $1.3655 \pm 0.0021$ & $1.2064 \pm 0.0021$ & 9.957 & \cite{2012ApJ...745..109L} \\
J1906+0746 & $1.322 \pm 0.011$ & $1.291 \pm 0.011$ & 0.166 & \cite{2015ApJ...798..118V} \\
J1829+2456 & $1.306 \pm 0.004$ & $1.299 \pm 0.004$ & 1.176 & \cite{2021MNRAS.500.4620H} \\
\hline
\end{tabular}
\label{tab:ns_masses}
\end{table*}

\section{Combined Mass Distribution Inference}\label{sec:combined}

\begin{figure}
    \centering
    \includegraphics[width=0.45\linewidth]{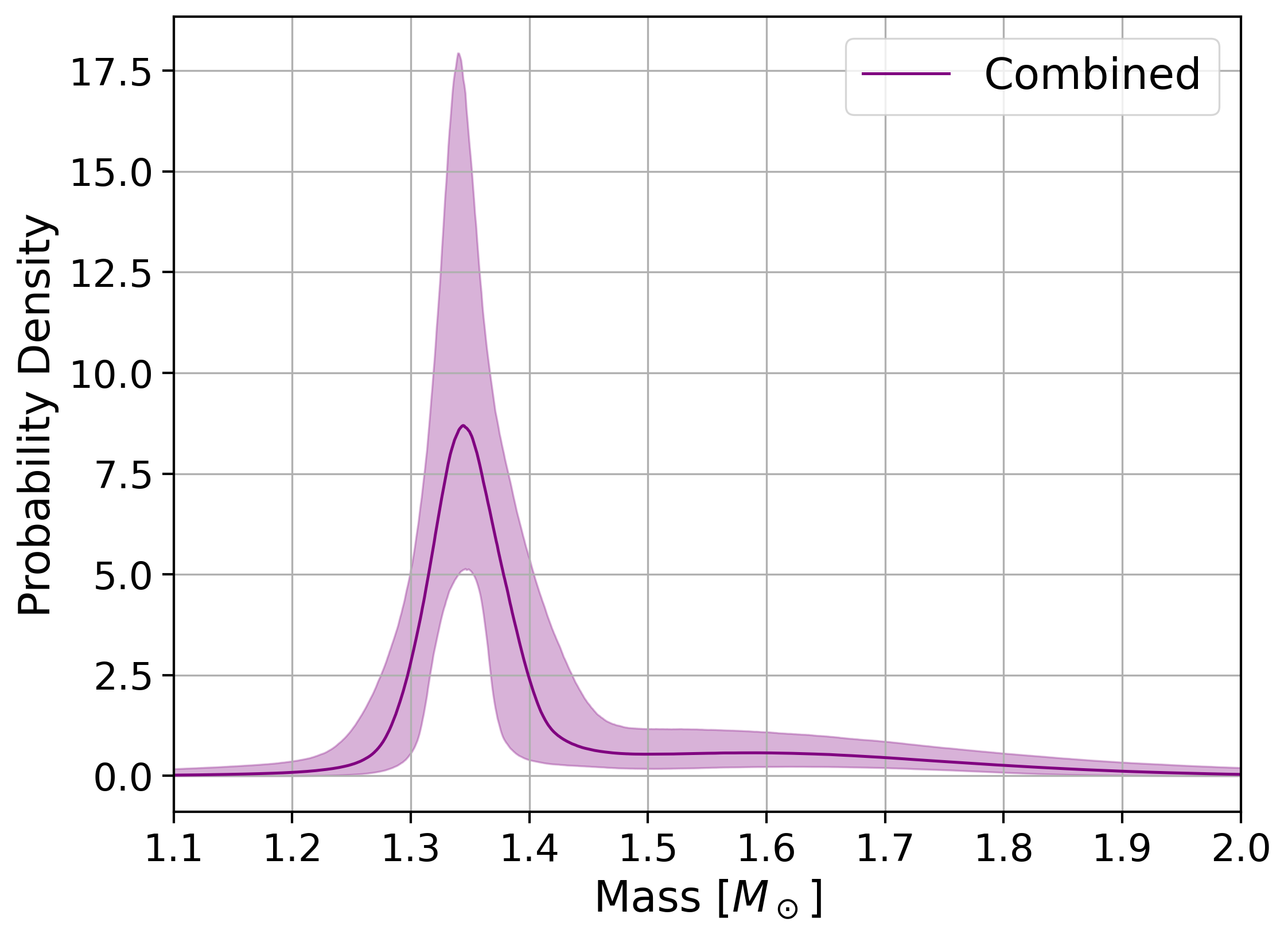}
    \includegraphics[width=0.45\linewidth]{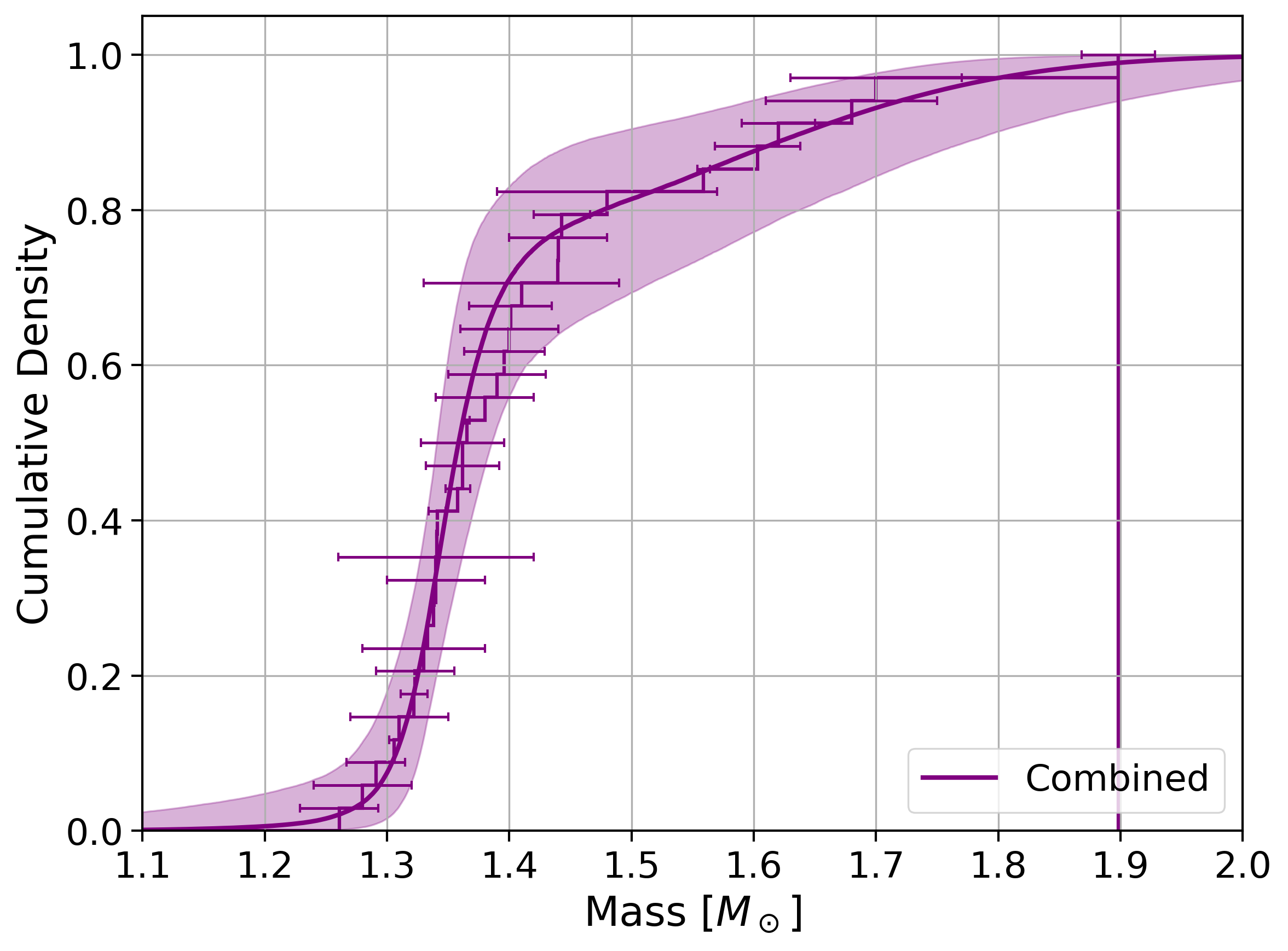}
    \caption{\textit{Left}: Posterior predictive distributions for the combined dataset of the \textit{Gaia} NS and recycled DNS populations. The bands show the 90\% credible regions. The $1.3 \text{ M}_\odot$ peak is well defined and there is clear support for a long tail to masses $>1.5 \text{ M}_\odot$, which is beyond the masses accretion could have achieved. \textit{Right:} Cumulative posterior predictive distributions shown with all combined data overplotted with errorbars. }
    \label{fig:massdist_combined}
\end{figure}

Our results suggest that both the \textit{Gaia} and first-born DNS populations have masses that are drawn from the same underlying astrophysical distribution. Here, we combine the two datasets to fit a single mass distribution, seen in Figure \ref{fig:massdist_combined}.

\begin{figure}
    \centering
    \includegraphics[width=1\linewidth]{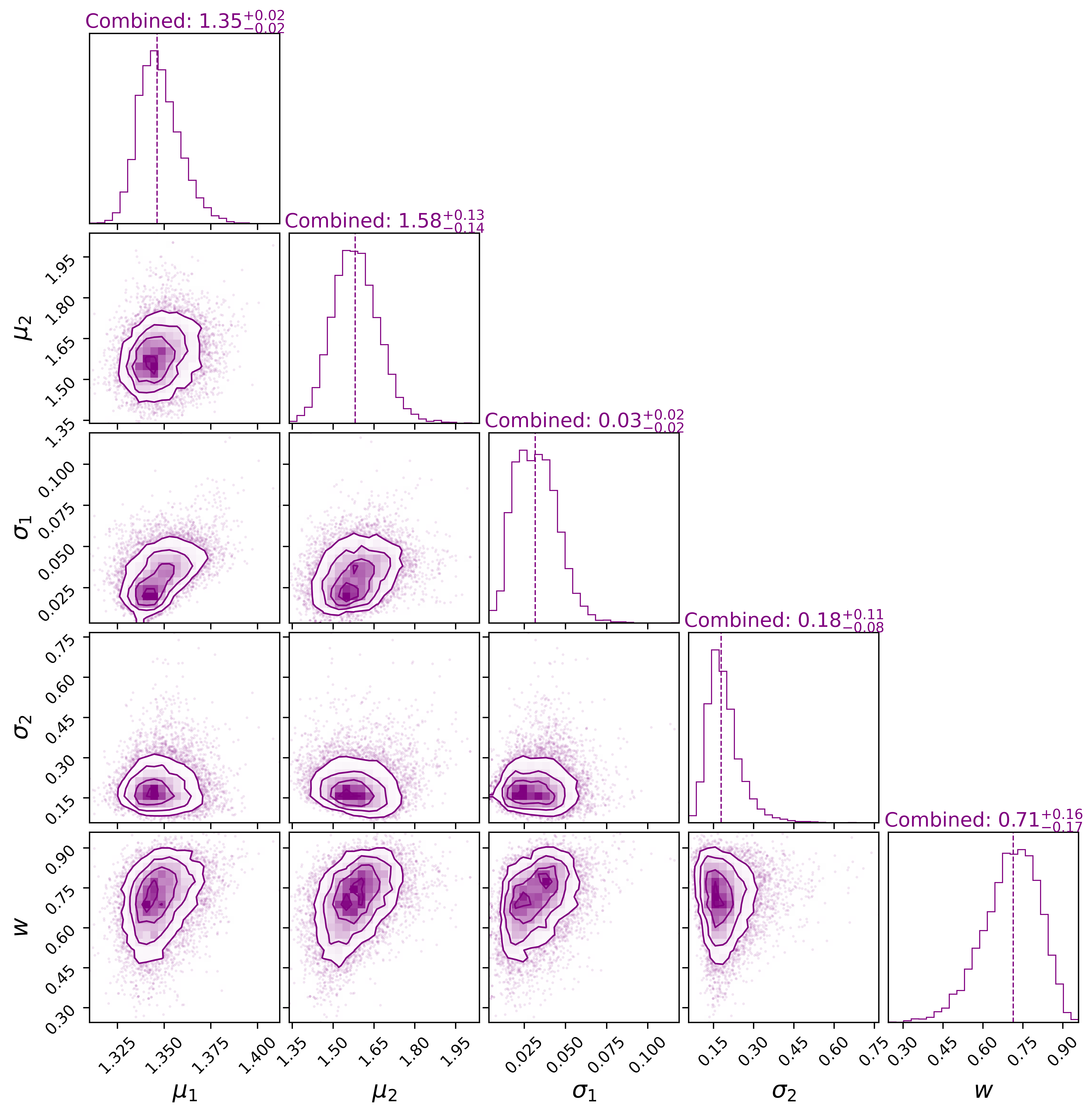}
    \caption{A corner plot showing the posteriors of 10,000 samples of the double Gaussian model for the combination of the \textit{Gaia} NS and recycled DNS populations. The numeric values shown are the median and 90\% credible limits. The dotted lines indicate the median values. These posteriors have more precise ranges than the separate fits (see Figure \ref{fig:corner}). The weight parameter is constrained away from the single Gaussian values of $w=0,1$.}
    \label{fig:corner_combined}
\end{figure}

In Figure \ref{fig:corner_combined}, it is seen that the model parameters are more tightly constrained than the separate fits in Figure \ref{fig:corner} due to the large dataset. The combined dataset more strongly excludes the single Gaussian model as seen in the marginal plot for the weight parameter, which is constrained away from $w=0,1$. Table \ref{t:hdi} also shows that the HDI no longer includes $w=1$ in the combined data set with 90\% coverage over $w=$[0.54,0.88]. Consequently, the island of support in $\mu_2$ near $\mu_1\sim 1.3 \text{ M}_\odot$ in Figure \ref{fig:corner} is not present in this combined inference. 

\begin{table}[ht!]
\centering 
\caption{The 90\% HDIs of the double Gaussian model parameters, for the fit to the \textit{Gaia} NS data, the recycled DNS data, and the combination of both.}
\label{t:hdi}
\begin{tabular}{cccc}
\toprule
Parameter & \textit{Gaia} NS & Recycled DNS & Combined \\
\midrule\midrule
  $\mu_1$          &[1.32, 1.40] & [1.29, 1.40] & [1.33, 1.37]\\
  $\sigma_1$       &[0.009, 0.12] & [0.003, 0.24] & [0.01, 0.05]\\
  $\mu_2$          &[1.36, 1.85] & [1.33, 1.69] & [1.44, 1.71]\\
  $\sigma_2$       &[0.03, 0.45] & [0.001, 0.47] & [0.09, 0.28]\\
  $w$              &[0.47, 0.97] & [0.33, 1.00] & [0.54, 0.88]\\
\midrule\midrule
\end{tabular}
\end{table}

This combined two-component mass distribution has a strong, first peak at $\mu_1<1.37$ with width $\sigma_1 <0.05$. This is notable as the small dispersion value has been considered a unique feature of the recycled DNSs and indicative of a specific and precise formation mechanism; however this is hard to explain with current understandings of fallback physics \citep{2012ApJ...757...55O}. Whatever physical origin underlies this narrow first peak, it must be common to first-born NSs in drastically different systems -- Galactic DNS and wide binaries with solar-type companions -- suggesting universality. 

Figure \ref{fig:massdist_combined} shows the fit CDF of the combined data set alongside the data points. Above 1.4(1.8)$ \text{ M}_\odot$ the 90\% credible region contains between 17--44(5--23)\% probability mass.

\clearpage
\bibliography{bibtex}{}

@ARTICLE{2025NatAs...9..552Y,
       author = {{You}, Zhi-Qiang and {Zhu}, Xingjiang and {Liu}, Xiaojin and {M{\"u}ller}, Bernhard and {Heger}, Alexander and {Stevenson}, Simon and {Thrane}, Eric and {Chen}, Zu-Cheng and {Sun}, Ling and {Lasky}, Paul and {Galloway}, Duncan K. and {Hobbs}, George and {Manchester}, Richard N. and {Gao}, He and {Zhu}, Zong-Hong},
        title = "{Determination of the birth-mass function of neutron stars from observations}",
      journal = {Nature Astronomy},
         year = 2025,
        month = apr,
       volume = {9},
        pages = {552-563},
          doi = {10.1038/s41550-025-02487-w},
archivePrefix = {arXiv},
       eprint = {2412.05524},
 primaryClass = {astro-ph.HE},
       adsurl = { https://ui.adsabs.harvard.edu/abs/2025NatAs...9..552Y}
}

@ARTICLE{2020RNAAS...4...65F,
       author = {{Farr}, Will M. and {Chatziioannou}, Katerina},
        title = "{A Population-Informed Mass Estimate for Pulsar J0740+6620}",
      journal = {Research Notes of the American Astronomical Society},
         year = 2020,
        month = may,
       volume = {4},
       number = {5},
          eid = {65},
        pages = {65},
          doi = {10.3847/2515-5172/ab9088},
archivePrefix = {arXiv},
       eprint = {2005.00032},
 primaryClass = {astro-ph.GA},
       adsurl = {https://ui.adsabs.harvard.edu/abs/2020RNAAS...4...65F}
}

@ARTICLE{2019ApJ...876...18F,
       author = {{Farrow}, Nicholas and {Zhu}, Xing-Jiang and {Thrane}, Eric},
        title = "{The Mass Distribution of Galactic Double Neutron Stars}",
      journal = {\apj},
         year = 2019,
        month = may,
       volume = {876},
       number = {1},
          eid = {18},
        pages = {18},
          doi = {10.3847/1538-4357/ab12e3},
archivePrefix = {arXiv},
       eprint = {1902.03300},
 primaryClass = {astro-ph.HE},
       adsurl = {https://ui.adsabs.harvard.edu/abs/2019ApJ...876...18F}
}

@ARTICLE{2017ApJ...846..170T,
       author = {{Tauris}, T.~M. and {Kramer}, M. and {Freire}, P.~C.~C. and {Wex}, N. and {Janka}, H. -T. and {Langer}, N. and {Podsiadlowski}, Ph. and {Bozzo}, E. and {Chaty}, S. and {Kruckow}, M.~U. and {van den Heuvel}, E.~P.~J. and {Antoniadis}, J. and {Breton}, R.~P. and {Champion}, D.~J.},
        title = "{Formation of Double Neutron Star Systems}",
      journal = {\apj},
         year = 2017,
        month = sep,
       volume = {846},
       number = {2},
          eid = {170},
        pages = {170},
          doi = {10.3847/1538-4357/aa7e89},
archivePrefix = {arXiv},
       eprint = {1706.09438},
 primaryClass = {astro-ph.HE},
       adsurl = {https://ui.adsabs.harvard.edu/abs/2017ApJ...846..170T}
}

@ARTICLE{2016ARA&A..54..401O,
       author = {{{\"O}zel}, Feryal and {Freire}, Paulo},
        title = "{Masses, Radii, and the Equation of State of Neutron Stars}",
      journal = {\araa},
         year = 2016,
        month = sep,
       volume = {54},
        pages = {401-440},
          doi = {10.1146/annurev-astro-081915-023322},
archivePrefix = {arXiv},
       eprint = {1603.02698},
 primaryClass = {astro-ph.HE},
       adsurl = {https://ui.adsabs.harvard.edu/abs/2016ARA&A..54..401O}
}

@ARTICLE{2012ApJ...757...55O,
       author = {{{\"O}zel}, Feryal and {Psaltis}, Dimitrios and {Narayan}, Ramesh and {Santos Villarreal}, Antonio},
        title = "{On the Mass Distribution and Birth Masses of Neutron Stars}",
      journal = {\apj},
         year = 2012,
        month = sep,
       volume = {757},
       number = {1},
          eid = {55},
        pages = {55},
          doi = {10.1088/0004-637X/757/1/55},
archivePrefix = {arXiv},
       eprint = {1201.1006},
 primaryClass = {astro-ph.HE},
       adsurl = {https://ui.adsabs.harvard.edu/abs/2012ApJ...757...55O}
}

@ARTICLE{2014A&A...561A..11V,
       author = {{Verbunt}, Frank and {Freire}, Paulo C.~C.},
        title = "{On the disruption of pulsar and X-ray binar ies in globular clusters}",
      journal = {\aap},
         year = 2014,
        month = jan,
       volume = {561},
          eid = {A11},
        pages = {A11},
          doi = {10.1051/0004-6361/201321177},
archivePrefix = {arXiv},
       eprint = {1310.4669},
 primaryClass = {astro-ph.SR},
       adsurl = {https://ui.adsabs.harvard.edu/abs/2014A&A...561A..11V}
}

@ARTICLE{2011MNRAS.413..461O,
       author = {{Os{\l}owski}, S. and {Bulik}, T. and {Gondek-Rosi{\'n}ska}, D. and {Belczy{\'n}ski}, K.},
        title = "{Population synthesis of double neutron stars}",
      journal = {\mnras},
         year = 2011,
        month = may,
       volume = {413},
       number = {1},
        pages = {461-479},
          doi = {10.1111/j.1365-2966.2010.18147.x},
archivePrefix = {arXiv},
       eprint = {0903.3538},
 primaryClass = {astro-ph.GA},
       adsurl = {https://ui.adsabs.harvard.edu/abs/2011MNRAS.413..461O}
}

@ARTICLE{2018MNRAS.478.1377A,
       author = {{Alsing}, Justin and {Silva}, Hector O. and {Berti}, Emanuele},
        title = "{Evidence for a maximum mass cut-off in the neutron star mass distribution and constraints on the equation of state}",
      journal = {\mnras},
         year = 2018,
        month = jul,
       volume = {478},
       number = {1},
        pages = {1377-1391},
          doi = {10.1093/mnras/sty1065},
archivePrefix = {arXiv},
       eprint = {1709.07889},
 primaryClass = {astro-ph.HE},
       adsurl = {https://ui.adsabs.harvard.edu/abs/2018MNRAS.478.1377A}
}

@ARTICLE{2024OJAp....7E..58E,
       author = {{El-Badry}, Kareem and {Rix}, Hans-Walter and {Latham}, David W. and {Shahaf}, Sahar and {Mazeh}, Tsevi and {Bieryla}, Allyson and {Buchhave}, Lars A. and {Andrae}, Ren{\'e} and {Yamaguchi}, Natsuko and {Isaacson}, Howard and {Howard}, Andrew W. and {Savino}, Alessandro and {Ilyin}, Ilya V.},
        title = "{A population of neutron star candidates in wide orbits from Gaia astrometry}",
      journal = {The Open Journal of Astrophysics},
         year = 2024,
        month = jul,
       volume = {7},
          eid = {58},
        pages = {58},
          doi = {10.33232/001c.121261},
archivePrefix = {arXiv},
       eprint = {2405.00089},
 primaryClass = {astro-ph.SR},
       adsurl = {https://ui.adsabs.harvard.edu/abs/2024OJAp....7E..58E}
}

@ARTICLE{2010ApJ...719..722S,
       author = {{Schwab}, J. and {Podsiadlowski}, Ph. and {Rappaport}, S.},
        title = "{Further Evidence for the Bimodal Distribution of Neutron-star Masses}",
      journal = {\apj},
         year = 2010,
        month = aug,
       volume = {719},
       number = {1},
        pages = {722-727},
          doi = {10.1088/0004-637X/719/1/722},
archivePrefix = {arXiv},
       eprint = {1006.4584},
 primaryClass = {astro-ph.HE},
       adsurl = {https://ui.adsabs.harvard.edu/abs/2010ApJ...719..722S}
}

@ARTICLE{2021ApJ...922..158L,
       author = {{Li}, Zhenwei and {Chen}, Xuefei and {Chen}, Hai-Liang and {Han}, Zhanwen},
        title = "{The Maximum Accreted Mass of Recycled Pulsars}",
      journal = {\apj},
         year = 2021,
        month = dec,
       volume = {922},
       number = {2},
          eid = {158},
        pages = {158},
          doi = {10.3847/1538-4357/ac1b2e},
archivePrefix = {arXiv},
       eprint = {2108.02554},
 primaryClass = {astro-ph.HE},
       adsurl = {https://ui.adsabs.harvard.edu/abs/2021ApJ...922..158L}
}

@ARTICLE{2024NewAR..9801694E,
       author = {{El-Badry}, Kareem},
        title = "{Gaia's binary star renaissance}",
      journal = {\nar},
         year = 2024,
        month = jun,
       volume = {98},
          eid = {101694},
        pages = {101694},
          doi = {10.1016/j.newar.2024.101694},
archivePrefix = {arXiv},
       eprint = {2403.12146},
 primaryClass = {astro-ph.SR},
       adsurl = {https://ui.adsabs.harvard.edu/abs/2024NewAR..9801694E}
}

@ARTICLE{2018arXiv181009538B,
       author = {{Bingham}, Eli and {Chen}, Jonathan P. and {Jankowiak}, Martin and {Obermeyer}, Fritz and {Pradhan}, Neeraj and {Karaletsos}, Theofanis and {Singh}, Rohit and {Szerlip}, Paul and {Horsfall}, Paul and {Goodman}, Noah D.},
        title = "{Pyro: Deep Universal Probabilistic Programming}",
      journal = {arXiv e-prints},
         year = 2018,
        month = oct,
          eid = {arXiv:1810.09538},
        pages = {arXiv:1810.09538},
          doi = {10.48550/arXiv.1810.09538},
archivePrefix = {arXiv},
       eprint = {1810.09538},
 primaryClass = {cs.LG},
       adsurl = {https://ui.adsabs.harvard.edu/abs/2018arXiv181009538B}
}

@ARTICLE{2019arXiv191211554P,
       author = {{Phan}, Du and {Pradhan}, Neeraj and {Jankowiak}, Martin},
        title = "{Composable Effects for Flexible and Accelerated Probabilistic Programming in NumPyro}",
      journal = {arXiv e-prints},
         year = 2019,
        month = dec,
          eid = {arXiv:1912.11554},
        pages = {arXiv:1912.11554},
          doi = {10.48550/arXiv.1912.11554},
archivePrefix = {arXiv},
       eprint = {1912.11554},
 primaryClass = {stat.ML},
       adsurl = {https://ui.adsabs.harvard.edu/abs/2019arXiv191211554P}
}

@software{jax2018github,
  author = {James Bradbury and Roy Frostig and Peter Hawkins and Matthew James Johnson and Chris Leary and Dougal Maclaurin and George Necula and Adam Paszke and Jake Vander{P}las and Skye Wanderman-{M}ilne and Qiao Zhang},
  title = {{JAX}: composable transformations of {P}ython+{N}um{P}y programs},
  url = {http://github.com/google/jax},
  version = {0.3.13},
  year = {2018},
}

@ARTICLE{2011arXiv1111.4246H,
       author = {{Hoffman}, Matthew D. and {Gelman}, Andrew},
        title = "{The No-U-Turn Sampler: Adaptively Setting Path Lengths in Hamiltonian Monte Carlo}",
      journal = {arXiv e-prints},
         year = 2011,
        month = nov,
          eid = {arXiv:1111.4246},
        pages = {arXiv:1111.4246},
          doi = {10.48550/arXiv.1111.4246},
archivePrefix = {arXiv},
       eprint = {1111.4246},
 primaryClass = {stat.CO},
       adsurl = {https://ui.adsabs.harvard.edu/abs/2011arXiv1111.4246H}
}

@ARTICLE{2011MNRAS.414.1427V,
       author = {{Valentim}, R. and {Rangel}, E. and {Horvath}, J.~E.},
        title = "{On the mass distribution of neutron stars}",
      journal = {\mnras},
         year = 2011,
        month = jun,
       volume = {414},
       number = {2},
        pages = {1427-1431},
          doi = {10.1111/j.1365-2966.2011.18477.x},
archivePrefix = {arXiv},
       eprint = {1101.4872},
 primaryClass = {astro-ph.SR},
       adsurl = {https://ui.adsabs.harvard.edu/abs/2011MNRAS.414.1427V}
}

@ARTICLE{2013arXiv1309.6635K,
       author = {{Kiziltan}, Bulent and {Kottas}, Athanasios and {De Yoreo}, Maria and {Thorsett}, Stephen E.},
        title = "{The Neutron Star Mass Distribution}",
      journal = {arXiv e-prints},
         year = 2013,
        month = sep,
          eid = {arXiv:1309.6635},
        pages = {arXiv:1309.6635},
          doi = {10.48550/arXiv.1309.6635},
archivePrefix = {arXiv},
       eprint = {1309.6635},
 primaryClass = {astro-ph.SR},
       adsurl = {https://ui.adsabs.harvard.edu/abs/2013arXiv1309.6635K}
}

@ARTICLE{2016arXiv160501665A,
       author = {{Antoniadis}, John and {Tauris}, Thomas M. and {Ozel}, Feryal and {Barr}, Ewan and {Champion}, David J. and {Freire}, Paulo C.~C.},
        title = "{The millisecond pulsar mass distribution: Evidence for bimodality and constraints on the maximum neutron star mass}",
      journal = {arXiv e-prints},
         year = 2016,
        month = may,
          eid = {arXiv:1605.01665},
        pages = {arXiv:1605.01665},
          doi = {10.48550/arXiv.1605.01665},
archivePrefix = {arXiv},
       eprint = {1605.01665},
 primaryClass = {astro-ph.HE},
       adsurl = {https://ui.adsabs.harvard.edu/abs/2016arXiv160501665A}
}

@ARTICLE{1996ApJ...457..834T,
       author = {{Timmes}, F.~X. and {Woosley}, S.~E. and {Weaver}, Thomas A.},
        title = "{The Neutron Star and Black Hole Initial Mass Function}",
      journal = {\apj},
         year = 1996,
        month = feb,
       volume = {457},
        pages = {834},
          doi = {10.1086/176778},
archivePrefix = {arXiv},
       eprint = {astro-ph/9510136},
 primaryClass = {astro-ph},
       adsurl = {https://ui.adsabs.harvard.edu/abs/1996ApJ...457..834T}
}

@BOOK{2004hpa..book.....L,
       author = {{Lorimer}, D.~R. and {Kramer}, M.},
        title = "{Handbook of Pulsar Astronomy}",
         year = 2004,
       volume = {4},
        publisher = {Cambridge University Press},
       adsurl = {https://ui.adsabs.harvard.edu/abs/2004hpa..book.....L}
}

@ARTICLE{2004ApJ...612.1044P,
       author = {{Podsiadlowski}, Ph. and {Langer}, N. and {Poelarends}, A.~J.~T. and {Rappaport}, S. and {Heger}, A. and {Pfahl}, E.},
        title = "{The Effects of Binary Evolution on the Dynamics of Core Collapse and Neutron Star Kicks}",
      journal = {\apj},
         year = 2004,
        month = sep,
       volume = {612},
       number = {2},
        pages = {1044-1051},
          doi = {10.1086/421713},
archivePrefix = {arXiv},
       eprint = {astro-ph/0309588},
 primaryClass = {astro-ph},
       adsurl = {https://ui.adsabs.harvard.edu/abs/2004ApJ...612.1044P}
}

@ARTICLE{2024ApJ...966...26T,
       author = {{Tan}, Chia Min and {Fonseca}, Emmanuel and {Crowter}, Kathryn and {Dong}, Fengqiu Adam and {Kaspi}, Victoria M. and {Masui}, Kiyoshi W. and {McKee}, James W. and {Meyers}, Bradley W. and {Ransom}, Scott M. and {Stairs}, Ingrid H.},
        title = "{High-cadence Timing of Binary Pulsars with CHIME}",
      journal = {\apj},
         year = 2024,
        month = may,
       volume = {966},
       number = {1},
          eid = {26},
        pages = {26},
          doi = {10.3847/1538-4357/ad28b2},
archivePrefix = {arXiv},
       eprint = {2402.08188},
 primaryClass = {astro-ph.HE},
       adsurl = {https://ui.adsabs.harvard.edu/abs/2024ApJ...966...26T}
}

@INPROCEEDINGS{2015aska.confE..40K,
       author = {{Keane}, E. and {Bhattacharyya}, B. and {Kramer}, M. and {Stappers}, B. and {Keane}, E.~F. and {Bhattacharyya}, B. and {Kramer}, M. and {Stappers}, B.~W. and {Bates}, S.~D. and {Burgay}, M. and {Chatterjee}, S. and {Champion}, D.~J. and {Eatough}, R.~P. and {Hessels}, J.~W.~T. and {Janssen}, G. and {Lee}, K.~J. and {van Leeuwen}, J. and {Margueron}, J. and {Oertel}, M. and {Possenti}, A. and {Ransom}, S. and {Theureau}, G. and {Torne}, P.},
        title = "{A Cosmic Census of Radio Pulsars with the SKA}",
    booktitle = {Advancing Astrophysics with the Square Kilometre Array (AASKA14)},
         year = 2015,
        month = apr,
          eid = {40},
        pages = {40},
          doi = {10.22323/1.215.0040},
archivePrefix = {arXiv},
       eprint = {1501.00056},
 primaryClass = {astro-ph.IM},
       adsurl = {https://ui.adsabs.harvard.edu/abs/2015aska.confE..40K}
}

@ARTICLE{2019MNRAS.483.2615K,
       author = {{Kyutoku}, Koutarou and {Nishino}, Yuki and {Seto}, Naoki},
        title = "{How to detect the shortest period binary pulsars in the era of LISA}",
      journal = {\mnras},
         year = 2019,
        month = feb,
       volume = {483},
       number = {2},
        pages = {2615-2620},
          doi = {10.1093/mnras/sty3322},
archivePrefix = {arXiv},
       eprint = {1812.02177},
 primaryClass = {astro-ph.HE},
       adsurl = {https://ui.adsabs.harvard.edu/abs/2019MNRAS.483.2615K}
}

@ARTICLE{2025RAA....25a4003W,
       author = {{Wang}, P.~F. and {Han}, J.~L. and {Yang}, Z.~L. and {Wang}, T. and {Wang}, C. and {Su}, W.~Q. and {Xu}, J. and {Zhou}, D.~J. and {Yan}, Yi and {Jing}, W.~C. and {Cai}, N.~N. and {Yuan}, J.~P. and {Xu}, R.~X. and {Wang}, H.~G. and {You}, X.~P.},
        title = "{The FAST Galactic Plane Pulsar Snapshot Survey. VIII. 116 Binary Pulsars}",
      journal = {Research in Astronomy and Astrophysics},
         year = 2025,
        month = jan,
       volume = {25},
       number = {1},
          eid = {014003},
        pages = {014003},
          doi = {10.1088/1674-4527/ada3b8},
archivePrefix = {arXiv},
       eprint = {2412.03062},
 primaryClass = {astro-ph.HE},
       adsurl = {https://ui.adsabs.harvard.edu/abs/2025RAA....25a4003W}
}

@ARTICLE{2012MNRAS.425.1601T,
       author = {{Tauris}, T.~M. and {Langer}, N. and {Kramer}, M.},
        title = "{Formation of millisecond pulsars with CO white dwarf companions - II. Accretion, spin-up, true ages and comparison to MSPs with He white dwarf companions}",
      journal = {\mnras},
         year = 2012,
        month = sep,
       volume = {425},
       number = {3},
        pages = {1601-1627},
          doi = {10.1111/j.1365-2966.2012.21446.x},
archivePrefix = {arXiv},
       eprint = {1206.1862},
 primaryClass = {astro-ph.SR},
       adsurl = {https://ui.adsabs.harvard.edu/abs/2012MNRAS.425.1601T}
}

@ARTICLE{2015MNRAS.451.2123T,
       author = {{Tauris}, Thomas M. and {Langer}, Norbert and {Podsiadlowski}, Philipp},
        title = "{Ultra-stripped supernovae: progenitors and fate}",
      journal = {\mnras},
         year = 2015,
        month = aug,
       volume = {451},
       number = {2},
        pages = {2123-2144},
          doi = {10.1093/mnras/stv990},
archivePrefix = {arXiv},
       eprint = {1505.00270},
 primaryClass = {astro-ph.SR},
       adsurl = {https://ui.adsabs.harvard.edu/abs/2015MNRAS.451.2123T}
}

@ARTICLE{1991ApJ...374L..41P,
       author = {{Prince}, T.~A. and {Anderson}, S.~B. and {Kulkarni}, S.~R. and {Wolszczan}, A.},
        title = "{Timing Observations of the 8 Hour Binary Pulsar 2127+11C in the Globular Cluster M15}",
      journal = {\apjl},
         year = 1991,
        month = jun,
       volume = {374},
        pages = {L41},
          doi = {10.1086/186067},
       adsurl = {https://ui.adsabs.harvard.edu/abs/1991ApJ...374L..41P}
}

@ARTICLE{2012ApJ...745..109L,
       author = {{Lynch}, Ryan S. and {Freire}, Paulo C.~C. and {Ransom}, Scott M. and {Jacoby}, Bryan A.},
        title = "{The Timing of Nine Globular Cluster Pulsars}",
      journal = {\apj},
         year = 2012,
        month = feb,
       volume = {745},
       number = {2},
          eid = {109},
        pages = {109},
          doi = {10.1088/0004-637X/745/2/109},
archivePrefix = {arXiv},
       eprint = {1112.2612},
 primaryClass = {astro-ph.HE},
       adsurl = {https://ui.adsabs.harvard.edu/abs/2012ApJ...745..109L}
}

@ARTICLE{2014MNRAS.443.2183F,
       author = {{Ferdman}, R.~D. and {Stairs}, I.~H. and {Kramer}, M. and {Janssen}, G.~H. and {Bassa}, C.~G. and {Stappers}, B.~W. and {Demorest}, P.~B. and {Cognard}, I. and {Desvignes}, G. and {Theureau}, G. and {Burgay}, M. and {Lyne}, A.~G. and {Manchester}, R.~N. and {Possenti}, A.},
        title = "{PSR J1756-2251: a pulsar with a low-mass neutron star companion}",
      journal = {\mnras},
         year = 2014,
        month = sep,
       volume = {443},
       number = {3},
        pages = {2183-2196},
          doi = {10.1093/mnras/stu1223},
archivePrefix = {arXiv},
       eprint = {1406.5507},
 primaryClass = {astro-ph.SR},
       adsurl = {https://ui.adsabs.harvard.edu/abs/2014MNRAS.443.2183F}
}

@ARTICLE{2006Sci...314...97K,
       author = {{Kramer}, M. and {Stairs}, I.~H. and {Manchester}, R.~N. and {McLaughlin}, M.~A. and {Lyne}, A.~G. and {Ferdman}, R.~D. and {Burgay}, M. and {Lorimer}, D.~R. and {Possenti}, A. and {D'Amico}, N. and {Sarkissian}, J.~M. and {Hobbs}, G.~B. and {Reynolds}, J.~E. and {Freire}, P.~C.~C. and {Camilo}, F.},
        title = "{Tests of General Relativity from Timing the Double Pulsar}",
      journal = {Science},
         year = 2006,
        month = oct,
       volume = {314},
       number = {5796},
        pages = {97-102},
          doi = {10.1126/science.1132305},
archivePrefix = {arXiv},
       eprint = {astro-ph/0609417},
 primaryClass = {astro-ph},
       adsurl = {https://ui.adsabs.harvard.edu/abs/2006Sci...314...97K}
}

@ARTICLE{2015ApJ...798..118V,
       author = {{van Leeuwen}, J. and {Kasian}, L. and {Stairs}, I.~H. and {Lorimer}, D.~R. and {Camilo}, F. and {Chatterjee}, S. and {Cognard}, I. and {Desvignes}, G. and {Freire}, P.~C.~C. and {Janssen}, G.~H. and {Kramer}, M. and {Lyne}, A.~G. and {Nice}, D.~J. and {Ransom}, S.~M. and {Stappers}, B.~W. and {Weisberg}, J.~M.},
        title = "{The Binary Companion of Young, Relativistic Pulsar J1906+0746}",
      journal = {\apj},
         year = 2015,
        month = jan,
       volume = {798},
       number = {2},
          eid = {118},
        pages = {118},
          doi = {10.1088/0004-637X/798/2/118},
archivePrefix = {arXiv},
       eprint = {1411.1518},
 primaryClass = {astro-ph.SR},
       adsurl = {https://ui.adsabs.harvard.edu/abs/2015ApJ...798..118V}
}

@ARTICLE{2014ApJ...787...82F,
       author = {{Fonseca}, Emmanuel and {Stairs}, Ingrid H. and {Thorsett}, Stephen E.},
        title = "{A Comprehensive Study of Relativistic Gravity Using PSR B1534+12}",
      journal = {\apj},
         year = 2014,
        month = may,
       volume = {787},
       number = {1},
          eid = {82},
        pages = {82},
          doi = {10.1088/0004-637X/787/1/82},
archivePrefix = {arXiv},
       eprint = {1402.4836},
 primaryClass = {astro-ph.HE},
       adsurl = {https://ui.adsabs.harvard.edu/abs/2014ApJ...787...82F}
}

@ARTICLE{2006ApJ...644L.113J,
       author = {{Jacoby}, B.~A. and {Cameron}, P.~B. and {Jenet}, F.~A. and {Anderson}, S.~B. and {Murty}, R.~N. and {Kulkarni}, S.~R.},
        title = "{Measurement of Orbital Decay in the Double Neutron Star Binary PSR B2127+11C}",
      journal = {\apjl},
         year = 2006,
        month = jun,
       volume = {644},
       number = {2},
        pages = {L113-L116},
          doi = {10.1086/505742},
archivePrefix = {arXiv},
       eprint = {astro-ph/0605375},
 primaryClass = {astro-ph},
       adsurl = {https://ui.adsabs.harvard.edu/abs/2006ApJ...644L.113J}
}

@ARTICLE{2008AA...490..753J,
       author = {{Janssen}, G.~H. and {Stappers}, B.~W. and {Kramer}, M. and {Nice}, D.~J. and {Jessner}, A. and {Cognard}, I. and {Purver}, M.~B.},
        title = "{Multi-telescope timing of PSR J1518+4904}",
      journal = {\aap},
         year = 2008,
        month = nov,
       volume = {490},
       number = {2},
        pages = {753-761},
          doi = {10.1051/0004-6361:200810076},
archivePrefix = {arXiv},
       eprint = {0808.2292},
 primaryClass = {astro-ph},
       adsurl = {https://ui.adsabs.harvard.edu/abs/2008A&A...490..753J}
}

@ARTICLE{2018ApJ...859...93L,
       author = {{Lynch}, Ryan S. and {Swiggum}, Joseph K. and {Kondratiev}, Vlad I. and {Kaplan}, David L. and {Stovall}, Kevin and {Fonseca}, Emmanuel and {Roberts}, Mallory S.~E. and {Levin}, Lina and {DeCesar}, Megan E. and {Cui}, Bingyi and {Cenko}, S. Bradley and {Gatkine}, Pradip and {Archibald}, Anne M. and {Banaszak}, Shawn and {Biwer}, Christopher M. and {Boyles}, Jason and {Chawla}, Pragya and {Dartez}, Louis P. and {Day}, David and {Ford}, Anthony J. and {Flanigan}, Joseph and {Hessels}, Jason W.~T. and {Hinojosa}, Jesus and {Jenet}, Fredrick A. and {Karako-Argaman}, Chen and {Kaspi}, Victoria M. and {Leake}, Sean and {Lunsford}, Grady and {Martinez}, Jos{\'e} G. and {Mata}, Alberto and {McLaughlin}, Maura A. and {Noori}, Hind Al and {Ransom}, Scott M. and {Rohr}, Matthew D. and {Siemens}, Xavier and {Spiewak}, Ren{\'e}e and {Stairs}, Ingrid H. and {van Leeuwen}, Joeri and {Walker}, Arielle N. and {Wells}, Bradley L.},
        title = "{The Green Bank North Celestial Cap Pulsar Survey. III. 45 New Pulsar Timing Solutions}",
      journal = {\apj},
         year = 2018,
        month = jun,
       volume = {859},
       number = {2},
          eid = {93},
        pages = {93},
          doi = {10.3847/1538-4357/aabf8a},
archivePrefix = {arXiv},
       eprint = {1805.04951},
 primaryClass = {astro-ph.HE},
       adsurl = {https://ui.adsabs.harvard.edu/abs/2018ApJ...859...93L}
}

@ARTICLE{2010ApJ...722.1030W,
       author = {{Weisberg}, J.~M. and {Nice}, D.~J. and {Taylor}, J.~H.},
        title = "{Timing Measurements of the Relativistic Binary Pulsar PSR B1913+16}",
      journal = {\apj},
         year = 2010,
        month = oct,
       volume = {722},
       number = {2},
        pages = {1030-1034},
          doi = {10.1088/0004-637X/722/2/1030},
archivePrefix = {arXiv},
       eprint = {1011.0718},
 primaryClass = {astro-ph.GA},
       adsurl = {https://ui.adsabs.harvard.edu/abs/2010ApJ...722.1030W}
}

@ARTICLE{2015ApJ...812..143M,
       author = {{Martinez}, J.~G. and {Stovall}, K. and {Freire}, P.~C.~C. and {Deneva}, J.~S. and {Jenet}, F.~A. and {McLaughlin}, M.~A. and {Bagchi}, M. and {Bates}, S.~D. and {Ridolfi}, A.},
        title = "{Pulsar J0453+1559: A Double Neutron Star System with a Large Mass Asymmetry}",
      journal = {\apj},
         year = 2015,
        month = oct,
       volume = {812},
       number = {2},
          eid = {143},
        pages = {143},
          doi = {10.1088/0004-637X/812/2/143},
archivePrefix = {arXiv},
       eprint = {1509.08805},
 primaryClass = {astro-ph.HE},
       adsurl = {https://ui.adsabs.harvard.edu/abs/2015ApJ...812..143M}
}

@ARTICLE{2021MNRAS.500.4620H,
       author = {{Haniewicz}, H.~T. and {Ferdman}, R.~D. and {Freire}, P.~C.~C. and {Champion}, D.~J. and {Bunting}, K.~A. and {Lorimer}, D.~R. and {McLaughlin}, M.~A.},
        title = "{Precise mass measurements for the double neutron star system J1829+2456}",
      journal = {\mnras},
         year = 2021,
        month = feb,
       volume = {500},
       number = {4},
        pages = {4620-4627},
          doi = {10.1093/mnras/staa3466},
archivePrefix = {arXiv},
       eprint = {2007.07565},
 primaryClass = {astro-ph.SR},
       adsurl = {https://ui.adsabs.harvard.edu/abs/2021MNRAS.500.4620H}
}

@ARTICLE{2005AJ....129.1993M,
       author = {{Manchester}, R.~N. and {Hobbs}, G.~B. and {Teoh}, A. and {Hobbs}, M.},
        title = "{The Australia Telescope National Facility Pulsar Catalogue}",
      journal = {\aj},
         year = 2005,
        month = apr,
       volume = {129},
       number = {4},
        pages = {1993-2006},
          doi = {10.1086/428488},
archivePrefix = {arXiv},
       eprint = {astro-ph/0412641},
 primaryClass = {astro-ph},
       adsurl = {https://ui.adsabs.harvard.edu/abs/2005AJ....129.1993M}
}

@ARTICLE{2024MNRAS.530.1506S,
       author = {{Su}, W.~Q. and {Han}, J.~L. and {Yang}, Z.~L. and {Wang}, P.~F. and {Yuan}, J.~P. and {Wang}, C. and {Zhou}, D.~J. and {Wang}, T. and {Yan}, Y. and {Jing}, W.~C. and {Cai}, N.~N. and {Xie}, L. and {Xu}, J. and {Wang}, H.~G. and {Xu}, R.~X. and {You}, X.~P.},
        title = "{The FAST Galactic Plane Pulsar Snapshot Survey - V. PSR J1901+0658 in a double neutron star system}",
      journal = {\mnras},
         year = 2024,
        month = may,
       volume = {530},
       number = {2},
        pages = {1506-1511},
          doi = {10.1093/mnras/stae888},
archivePrefix = {arXiv},
       eprint = {2403.11635},
 primaryClass = {astro-ph.HE},
       adsurl = {https://ui.adsabs.harvard.edu/abs/2024MNRAS.530.1506S}
}

@ARTICLE{2021ApJ...921L..25L,
       author = {{Landry}, Philippe and {Read}, Jocelyn S.},
        title = "{The Mass Distribution of Neutron Stars in Gravitational-wave Binaries}",
      journal = {\apjl},
         year = 2021,
        month = nov,
       volume = {921},
       number = {2},
          eid = {L25},
        pages = {L25},
          doi = {10.3847/2041-8213/ac2f3e},
archivePrefix = {arXiv},
       eprint = {2107.04559},
 primaryClass = {astro-ph.HE},
       adsurl = {https://ui.adsabs.harvard.edu/abs/2021ApJ...921L..25L}
}

@ARTICLE{2020MNRAS.492.3061L,
       author = {{Lau}, Mike Y.~M. and {Mandel}, Ilya and {Vigna-G{\'o}mez}, Alejandro and {Neijssel}, Coenraad J. and {Stevenson}, Simon and {Sesana}, Alberto},
        title = "{Detecting double neutron stars with LISA}",
      journal = {\mnras},
         year = 2020,
        month = mar,
       volume = {492},
       number = {3},
        pages = {3061-3072},
          doi = {10.1093/mnras/staa002},
archivePrefix = {arXiv},
       eprint = {1910.12422},
 primaryClass = {astro-ph.HE},
       adsurl = {https://ui.adsabs.harvard.edu/abs/2020MNRAS.492.3061L}
}

@article{10.1214/aoms/1177729694,
author = {S. Kullback and R. A. Leibler},
title = {{On Information and Sufficiency}},
volume = {22},
journal = {The Annals of Mathematical Statistics},
number = {1},
publisher = {Institute of Mathematical Statistics},
pages = {79 -- 86},
year = {1951},
doi = {10.1214/aoms/1177729694},
URL = {https://doi.org/10.1214/aoms/1177729694}
}

@ARTICLE{61115,
  author={Lin, J.},
  journal={IEEE Transactions on Information Theory}, 
  title={Divergence measures based on the Shannon entropy}, 
  year={1991},
  volume={37},
  number={1},
  pages={145-151},
  doi={10.1109/18.61115}}

@Inbook{Villani2009,
author="Villani, C{\'e}dric",
title="The Wasserstein distances",
bookTitle="Optimal Transport: Old and New",
year="2009",
publisher="Springer Berlin Heidelberg",
address="Berlin, Heidelberg",
pages="93--111",
isbn="978-3-540-71050-9",
doi="10.1007/978-3-540-71050-9_6",
url="https://doi.org/10.1007/978-3-540-71050-9_6"
}

@ARTICLE{2013ApJ...775...27C,
       author = {{Chen}, Hai-Liang and {Chen}, Xuefei and {Tauris}, Thomas M. and {Han}, Zhanwen},
        title = "{Formation of Black Widows and Redbacks{\textemdash}Two Distinct Populations of Eclipsing Binary Millisecond Pulsars}",
      journal = {\apj},
         year = 2013,
        month = sep,
       volume = {775},
       number = {1},
          eid = {27},
        pages = {27},
          doi = {10.1088/0004-637X/775/1/27},
archivePrefix = {arXiv},
       eprint = {1308.4107},
 primaryClass = {astro-ph.SR},
       adsurl = {https://ui.adsabs.harvard.edu/abs/2013ApJ...775...27C}
}

@INPROCEEDINGS{2013IAUS..291..127R,
       author = {{Roberts}, Mallory S.~E.},
        title = "{Surrounded by spiders! New black widows and redbacks in the Galactic field}",
    booktitle = {Neutron Stars and Pulsars: Challenges and Opportunities after 80 years},
         year = 2013,
       editor = {{van Leeuwen}, Joeri},
       series = {IAU Symposium},
       volume = {291},
        month = mar,
        pages = {127-132},
          doi = {10.1017/S174392131202337X},
archivePrefix = {arXiv},
       eprint = {1210.6903},
 primaryClass = {astro-ph.HE},
       adsurl = {https://ui.adsabs.harvard.edu/abs/2013IAUS..291..127R}
}

@ARTICLE{2020Natur.583..211F,
       author = {{Ferdman}, R.~D. and {Freire}, P.~C.~C. and {Perera}, B.~B.~P. and {Pol}, N. and {Camilo}, F. and {Chatterjee}, S. and {Cordes}, J.~M. and {Crawford}, F. and {Hessels}, J.~W.~T. and {Kaspi}, V.~M. and {McLaughlin}, M.~A. and {Parent}, E. and {Stairs}, I.~H. and {van Leeuwen}, J.},
        title = "{Asymmetric mass ratios for bright double neutron-star mergers}",
      journal = {\nat},
         year = 2020,
        month = jul,
       volume = {583},
       number = {7815},
        pages = {211-214},
          doi = {10.1038/s41586-020-2439-x},
archivePrefix = {arXiv},
       eprint = {2007.04175},
 primaryClass = {astro-ph.HE},
       adsurl = {https://ui.adsabs.harvard.edu/abs/2020Natur.583..211F}
}

@ARTICLE{2023mgm..conf.3774C,
       author = {{Cameron}, A.~D. and {Bailes}, M. and {Balakrishnan}, V. and {Champion}, D.~J. and {Freire}, P.~C.~C. and {Kramer}, M. and {Wex}, N. and {Johnston}, S. and {Lyne}, A.~G. and {Stappers}, B.~W. and {McLaughlin}, M.~A. and {Pol}, N. and {Wahl}, H. and {Ng}, C. and {Possenti}, A. and {Ridolfi}, A.},
        title = "{News and views regarding PSR J1757-1854, a highly-relativistic binary pulsar}",
      journal = {The Sixteenth Marcel Grossmann Meeting. On Recent Developments in Theoretical and Experimental General Relativity, Astrophysics, and Relativistic Field Theories},
         year = 2023,
        month = jul,
        pages = {3774-3784},
          doi = {10.1142/9789811269776_0312},
archivePrefix = {arXiv},
       eprint = {2203.15995},
 primaryClass = {astro-ph.HE},
       adsurl = {https://ui.adsabs.harvard.edu/abs/2023mgm..conf.3774C}
}

@article{harris2020array,
  title={Array programming with NumPy},
  author={Harris, Charles R and Millman, K Jarrod and Van Der Walt, St{\'e}fan J and others},
  journal={Nature},
  volume={585},
  number={7825},
  pages={357--362},
  year={2020},
  publisher={Nature Publishing Group}
}

@article{virtanen2020scipy,
  title={SciPy 1.0: fundamental algorithms for scientific computing in Python},
  author={Virtanen, Pauli and Gommers, Ralf and Oliphant, Travis E and others},
  journal={Nature methods},
  volume={17},
  number={3},
  pages={261--272},
  year={2020},
  publisher={Nature Publishing Group}
}

@article{hunter2007matplotlib,
  title={Matplotlib: A 2D graphics environment},
  author={Hunter, John D},
  journal={Computing in science \& engineering},
  volume={9},
  number={3},
  pages={90--95},
  year={2007},
  publisher={IEEE}
}

@inproceedings{mckinney2010data,
  title={Data structures for statistical computing in python},
  author={McKinney, Wes and others},
  booktitle={Proceedings of the 9th Python in Science Conference},
  volume={445},
  pages={51--56},
  year={2010}
}

@article{waskom2021seaborn,
  title={seaborn: statistical data visualization},
  author={Waskom, Michael L},
  journal={Journal of Open Source Software},
  volume={6},
  number={60},
  pages={3021},
  year={2021}
}

@article{kumar2019arviz,
  title={ArviZ a unified library for exploratory analysis of Bayesian models in Python},
  author={Kumar, Ravin and Carroll, Colin and Hartikainen, Ari and Martin, Osvaldo},
  journal={Journal of Open Source Software},
  volume={4},
  number={33},
  pages={1143},
  year={2019}
}

@article{foreman2016corner,
  title={corner.py: Scatterplot matrices in Python},
  author={Foreman-Mackey, Daniel},
  journal={Journal of Open Source Software},
  volume={1},
  number={2},
  pages={24},
  year={2016}
}

@ARTICLE{2019MNRAS.482.2234G,
       author = {{Giacobbo}, Nicola and {Mapelli}, Michela},
        title = "{The impact of electron-capture supernovae on merging double neutron stars}",
      journal = {\mnras},
         year = 2019,
        month = jan,
       volume = {482},
       number = {2},
        pages = {2234-2243},
          doi = {10.1093/mnras/sty2848},
archivePrefix = {arXiv},
       eprint = {1805.11100},
 primaryClass = {astro-ph.SR},
       adsurl = {https://ui.adsabs.harvard.edu/abs/2019MNRAS.482.2234G}
}

@ARTICLE{1980PASJ...32..303M,
       author = {{Miyaji}, S. and {Nomoto}, K. and {Yokoi}, K. and {Sugimoto}, D.},
        title = "{Supernova triggered by electron captures.}",
      journal = {\pasj},
         year = 1980,
        month = jan,
       volume = {32},
        pages = {303-329},
       adsurl = {https://ui.adsabs.harvard.edu/abs/1980PASJ...32..303M}
}

@ARTICLE{1984ApJ...277..791N,
       author = {{Nomoto}, K.},
        title = "{Evolution of 8-10 solar mass stars toward electron capture supernovae. I - Formation of electron-degenerate O + NE + MG cores.}",
      journal = {\apj},
         year = 1984,
        month = feb,
       volume = {277},
        pages = {791-805},
          doi = {10.1086/161749},
       adsurl = {https://ui.adsabs.harvard.edu/abs/1984ApJ...277..791N}
}

@ARTICLE{1987ApJ...322..206N,
       author = {{Nomoto}, Ken'ichi},
        title = "{Evolution of 8--10 M$_{sun}$ Stars toward Electron Capture Supernovae. II. Collapse of an O + NE + MG Core}",
      journal = {\apj},
         year = 1987,
        month = nov,
       volume = {322},
        pages = {206},
          doi = {10.1086/165716},
       adsurl = {https://ui.adsabs.harvard.edu/abs/1987ApJ...322..206N}
}

@ARTICLE{1996Natur.381..584K,
       author = {{Kaspi}, V.~M. and {Bailes}, M. and {Manchester}, R.~N. and {Stappers}, B.~W. and {Bell}, J.~F.},
        title = "{Evidence from a precessing pulsar orbit for a neutron-star birth kick}",
      journal = {\nat},
         year = 1996,
        month = jun,
       volume = {381},
       number = {6583},
        pages = {584-586},
          doi = {10.1038/381584a0},
       adsurl = {https://ui.adsabs.harvard.edu/abs/1996Natur.381..584K}
}

@ARTICLE{2011MNRAS.412L..63B,
       author = {{Bassa}, C.~G. and {Brisken}, W.~F. and {Nelemans}, G. and {Stairs}, I.~H. and {Stappers}, B.~W. and {Kramer}, M.},
        title = "{The binary companion of PSR J1740-3052}",
      journal = {\mnras},
         year = 2011,
        month = mar,
       volume = {412},
       number = {1},
        pages = {L63-L67},
          doi = {10.1111/j.1745-3933.2010.01006.x},
archivePrefix = {arXiv},
       eprint = {1012.5254},
 primaryClass = {astro-ph.HE},
       adsurl = {https://ui.adsabs.harvard.edu/abs/2011MNRAS.412L..63B}
}

@ARTICLE{2014MNRAS.437.3255S,
       author = {{Shannon}, R.~M. and {Johnston}, S. and {Manchester}, R.~N.},
        title = "{The kinematics and orbital dynamics of the PSR B1259-63/LS 2883 system from 23 yr of pulsar timing}",
      journal = {\mnras},
         year = 2014,
        month = feb,
       volume = {437},
       number = {4},
        pages = {3255-3264},
          doi = {10.1093/mnras/stt2123},
archivePrefix = {arXiv},
       eprint = {1311.0588},
 primaryClass = {astro-ph.SR},
       adsurl = {https://ui.adsabs.harvard.edu/abs/2014MNRAS.437.3255S}
}

@ARTICLE{2015MNRAS.451..581L,
       author = {{Lyne}, A.~G. and {Stappers}, B.~W. and {Keith}, M.~J. and {Ray}, P.~S. and {Kerr}, M. and {Camilo}, F. and {Johnson}, T.~J.},
        title = "{The binary nature of PSR J2032+4127}",
      journal = {\mnras},
         year = 2015,
        month = jul,
       volume = {451},
       number = {1},
        pages = {581-587},
          doi = {10.1093/mnras/stv236},
archivePrefix = {arXiv},
       eprint = {1502.01465},
 primaryClass = {astro-ph.HE},
       adsurl = {https://ui.adsabs.harvard.edu/abs/2015MNRAS.451..581L}
}

@ARTICLE{2023ApJ...943...57A,
       author = {{Andersen}, Bridget C. and {Fonseca}, Emmanuel and {McKee}, J.~W. and {Meyers}, B.~W. and {Luo}, Jing and {Tan}, C.~M. and {Stairs}, I.~H. and {Kaspi}, Victoria M. and {van Kerkwijk}, M.~H. and {Bhardwaj}, Mohit and {Boyle}, P.~J. and {Crowter}, Kathryn and {Demorest}, Paul B. and {Dong}, Fengqiu A. and {Good}, Deborah C. and {Kaczmarek}, Jane F. and {Leung}, Calvin and {Masui}, Kiyoshi W. and {Naidu}, Arun and {Ng}, Cherry and {Patel}, Chitrang and {Pearlman}, Aaron B. and {Pleunis}, Ziggy and {Rafiei-Ravandi}, Masoud and {Rahman}, Mubdi and {Ransom}, Scott M. and {Smith}, Kendrick M. and {Tendulkar}, Shriharsh P.},
        title = "{CHIME Discovery of a Binary Pulsar with a Massive Nondegenerate Companion}",
      journal = {\apj},
         year = 2023,
        month = jan,
       volume = {943},
       number = {1},
          eid = {57},
        pages = {57},
          doi = {10.3847/1538-4357/aca485},
archivePrefix = {arXiv},
       eprint = {2209.06895},
 primaryClass = {astro-ph.HE},
       adsurl = {https://ui.adsabs.harvard.edu/abs/2023ApJ...943...57A}
}

@ARTICLE{2024A&A...682A.178V,
       author = {{van der Wateren}, E. and {Bassa}, C.~G. and {Janssen}, G.~H. and {Yanes-Rizo}, I.~V. and {Casares}, J. and {Nelemans}, G. and {Stappers}, B.~W. and {Tan}, C.~M.},
        title = "{PSR J0210+5845: Ultra-wide binary pulsar with a B6 V main sequence star companion}",
      journal = {\aap},
         year = 2024,
        month = feb,
       volume = {682},
          eid = {A178},
        pages = {A178},
          doi = {10.1051/0004-6361/202348578},
archivePrefix = {arXiv},
       eprint = {2312.01892},
 primaryClass = {astro-ph.HE},
       adsurl = {https://ui.adsabs.harvard.edu/abs/2024A&A...682A.178V}
}

@ARTICLE{2011Ap&SS.332....1R,
       author = {{Reig}, Pablo},
        title = "{Be/X-ray binaries}",
      journal = {\apss},
         year = 2011,
        month = mar,
       volume = {332},
       number = {1},
        pages = {1-29},
          doi = {10.1007/s10509-010-0575-8},
archivePrefix = {arXiv},
       eprint = {1101.5036},
 primaryClass = {astro-ph.HE},
       adsurl = {https://ui.adsabs.harvard.edu/abs/2011Ap&SS.332....1R}
}

@ARTICLE{2024A&A...684A.124F,
       author = {{Fortin}, F. and {Kalsi}, A. and {Garc{\'\i}a}, F. and {Simaz-Bunzel}, A. and {Chaty}, S.},
        title = "{A catalogue of low-mass X-ray binaries in the Galaxy: From the INTEGRAL to the Gaia era}",
      journal = {\aap},
         year = 2024,
        month = apr,
       volume = {684},
          eid = {A124},
        pages = {A124},
          doi = {10.1051/0004-6361/202347908},
archivePrefix = {arXiv},
       eprint = {2401.11931},
 primaryClass = {astro-ph.HE},
       adsurl = {https://ui.adsabs.harvard.edu/abs/2024A&A...684A.124F}
}

@ARTICLE{2017PhRvL.119p1101A,
       author = {{Abbott}, B.~P. and {Abbott}, R. and {Abbott}, T.~D. and {Acernese}, F. and {Ackley}, K. and {Adams}, C. and {Adams}, T. and {Addesso}, P. and {Adhikari}, R.~X. and {Adya}, V.~B. and {Affeldt}, C. and {Afrough}, M. and {Agarwal}, B. and {Agathos}, M. and {Agatsuma}, K. and {Aggarwal}, N. and {Aguiar}, O.~D. and {Aiello}, L. and {Ain}, A. and {Ajith}, P. and {Allen}, B. and {Allen}, G. and {Allocca}, A. and {Altin}, P.~A. and {Amato}, A. and {Ananyeva}, A. and {Anderson}, S.~B. and {Anderson}, W.~G. and {Angelova}, S.~V. and {Antier}, S. and {Appert}, S. and {Arai}, K. and {Araya}, M.~C. and {Areeda}, J.~S. and {Arnaud}, N. and {Arun}, K.~G. and {Ascenzi}, S. and {Ashton}, G. and {Ast}, M. and {Aston}, S.~M. and {Astone}, P. and {Atallah}, D.~V. and {Aufmuth}, P. and {Aulbert}, C. and {AultONeal}, K. and {Austin}, C. and {Avila-Alvarez}, A. and {Babak}, S. and {Bacon}, P. and {Bader}, M.~K.~M. and {Bae}, S. and {Bailes}, M. and {Baker}, P.~T. and {Baldaccini}, F. and {Ballardin}, G. and {Ballmer}, S.~W. and {Banagiri}, S. and {Barayoga}, J.~C. and {Barclay}, S.~E. and {Barish}, B.~C. and {Barker}, D. and {Barkett}, K. and {Barone}, F. and {Barr}, B. and {Barsotti}, L. and {Barsuglia}, M. and {Barta}, D. and {Barthelmy}, S.~D. and {Bartlett}, J. and {Bartos}, I. and {Bassiri}, R. and {Basti}, A. and {Batch}, J.~C. and {Bawaj}, M. and {Bayley}, J.~C. and {Bazzan}, M. and {B{\'e}csy}, B. and {Beer}, C. and {Bejger}, M. and {Belahcene}, I. and {Bell}, A.~S. and {Berger}, B.~K. and {Bergmann}, G. and {Bernuzzi}, S. and {Bero}, J.~J. and {Berry}, C.~P.~L. and {Bersanetti}, D. and {Bertolini}, A. and {Betzwieser}, J. and {Bhagwat}, S. and {Bhandare}, R. and {Bilenko}, I.~A. and {Billingsley}, G. and {Billman}, C.~R. and {Birch}, J. and {Birney}, R. and {Birnholtz}, O. and {Biscans}, S. and {Biscoveanu}, S. and {Bisht}, A. and {Bitossi}, M. and {Biwer}, C. and {Bizouard}, M.~A. and {Blackburn}, J.~K. and {Blackman}, J. and {Blair}, C.~D. and {Blair}, D.~G. and {Blair}, R.~M. and {Bloemen}, S. and {Bock}, O. and {Bode}, N. and {Boer}, M. and {Bogaert}, G. and {Bohe}, A. and {Bondu}, F. and {Bonilla}, E. and {Bonnand}, R. and {Boom}, B.~A. and {Bork}, R. and {Boschi}, V. and {Bose}, S. and {Bossie}, K. and {Bouffanais}, Y. and {Bozzi}, A. and {Bradaschia}, C. and {Brady}, P.~R. and {Branchesi}, M. and {Brau}, J.~E. and {Briant}, T. and {Brillet}, A. and {Brinkmann}, M. and {Brisson}, V. and {Brockill}, P. and {Broida}, J.~E. and {Brooks}, A.~F. and {Brown}, D.~A. and {Brown}, D.~D. and {Brunett}, S. and {Buchanan}, C.~C. and {Buikema}, A. and {Bulik}, T. and {Bulten}, H.~J. and {Buonanno}, A. and {Buskulic}, D. and {Buy}, C. and {Byer}, R.~L. and {Cabero}, M. and {Cadonati}, L. and {Cagnoli}, G. and {Cahillane}, C. and {Calder{\'o}n Bustillo}, J. and {Callister}, T.~A. and {Calloni}, E. and {Camp}, J.~B. and {Canepa}, M. and {Canizares}, P. and {Cannon}, K.~C. and {Cao}, H. and {Cao}, J. and {Capano}, C.~D. and {Capocasa}, E. and {Carbognani}, F. and {Caride}, S. and {Carney}, M.~F. and {Carullo}, G. and {Casanueva Diaz}, J. and {Casentini}, C. and {Caudill}, S. and {Cavagli{\`a}}, M. and {Cavalier}, F. and {Cavalieri}, R. and {Cella}, G. and {Cepeda}, C.~B. and {Cerd{\'a}-Dur{\'a}n}, P. and {Cerretani}, G. and {Cesarini}, E. and {Chamberlin}, S.~J. and {Chan}, M. and {Chao}, S. and {Charlton}, P. and {Chase}, E. and {Chassande-Mottin}, E. and {Chatterjee}, D. and {Chatziioannou}, K. and {Cheeseboro}, B.~D. and {Chen}, H.~Y. and {Chen}, X. and {Chen}, Y. and {Cheng}, H. -P. and {Chia}, H. and {Chincarini}, A. and {Chiummo}, A. and {Chmiel}, T. and {Cho}, H.~S. and {Cho}, M. and {Chow}, J.~H. and {Christensen}, N. and {Chu}, Q. and {Chua}, A.~J.~K. and {Chua}, S.},
        title = "{GW170817: Observation of Gravitational Waves from a Binary Neutron Star Inspiral}",
      journal = {\prl},
         year = 2017,
        month = oct,
       volume = {119},
       number = {16},
          eid = {161101},
        pages = {161101},
          doi = {10.1103/PhysRevLett.119.161101},
archivePrefix = {arXiv},
       eprint = {1710.05832},
 primaryClass = {gr-qc},
       adsurl = {https://ui.adsabs.harvard.edu/abs/2017PhRvL.119p1101A}
}

@ARTICLE{2020ApJ...892L...3A,
       author = {{Abbott}, B.~P. and {Abbott}, R. and {Abbott}, T.~D. and {Abraham}, S. and {Acernese}, F. and {Ackley}, K. and {Adams}, C. and {Adhikari}, R.~X. and {Adya}, V.~B. and {Affeldt}, C. and {Agathos}, M. and {Agatsuma}, K. and {Aggarwal}, N. and {Aguiar}, O.~D. and {Aiello}, L. and {Ain}, A. and {Ajith}, P. and {Allen}, G. and {Allocca}, A. and {Aloy}, M.~A. and {Altin}, P.~A. and {Amato}, A. and {Anand}, S. and {Ananyeva}, A. and {Anderson}, S.~B. and {Anderson}, W.~G. and {Angelova}, S.~V. and {Antier}, S. and {Appert}, S. and {Arai}, K. and {Araya}, M.~C. and {Areeda}, J.~S. and {Ar{\`e}ne}, M. and {Arnaud}, N. and {Aronson}, S.~M. and {Arun}, K.~G. and {Ascenzi}, S. and {Ashton}, G. and {Aston}, S.~M. and {Astone}, P. and {Aubin}, F. and {Aufmuth}, P. and {AultONeal}, K. and {Austin}, C. and {Avendano}, V. and {Avila-Alvarez}, A. and {Babak}, S. and {Bacon}, P. and {Badaracco}, F. and {Bader}, M.~K.~M. and {Bae}, S. and {Baird}, J. and {Baker}, P.~T. and {Baldaccini}, F. and {Ballardin}, G. and {Ballmer}, S.~W. and {Bals}, A. and {Banagiri}, S. and {Barayoga}, J.~C. and {Barbieri}, C. and {Barclay}, S.~E. and {Barish}, B.~C. and {Barker}, D. and {Barkett}, K. and {Barnum}, S. and {Barone}, F. and {Barr}, B. and {Barsotti}, L. and {Barsuglia}, M. and {Barta}, D. and {Bartlett}, J. and {Bartos}, I. and {Bassiri}, R. and {Basti}, A. and {Bawaj}, M. and {Bayley}, J.~C. and {Baylor}, A.~C. and {Bazzan}, M. and {B{\'e}csy}, B. and {Bejger}, M. and {Belahcene}, I. and {Bell}, A.~S. and {Beniwal}, D. and {Benjamin}, M.~G. and {Berger}, B.~K. and {Bergmann}, G. and {Bernuzzi}, S. and {Berry}, C.~P.~L. and {Bersanetti}, D. and {Bertolini}, A. and {Betzwieser}, J. and {Bhandare}, R. and {Bidler}, J. and {Biggs}, E. and {Bilenko}, I.~A. and {Bilgili}, S.~A. and {Billingsley}, G. and {Birney}, R. and {Birnholtz}, O. and {Biscans}, S. and {Bischi}, M. and {Biscoveanu}, S. and {Bisht}, A. and {Bitossi}, M. and {Bizouard}, M.~A. and {Blackburn}, J.~K. and {Blackman}, J. and {Blair}, C.~D. and {Blair}, D.~G. and {Blair}, R.~M. and {Bloemen}, S. and {Bobba}, F. and {Bode}, N. and {Boer}, M. and {Boetzel}, Y. and {Bogaert}, G. and {Bondu}, F. and {Bonnand}, R. and {Booker}, P. and {Boom}, B.~A. and {Bork}, R. and {Boschi}, V. and {Bose}, S. and {Bossilkov}, V. and {Bosveld}, J. and {Bouffanais}, Y. and {Bozzi}, A. and {Bradaschia}, C. and {Brady}, P.~R. and {Bramley}, A. and {Branchesi}, M. and {Brau}, J.~E. and {Breschi}, M. and {Briant}, T. and {Briggs}, J.~H. and {Brighenti}, F. and {Brillet}, A. and {Brinkmann}, M. and {Brockill}, P. and {Brooks}, A.~F. and {Brooks}, J. and {Brown}, D.~D. and {Brunett}, S. and {Buikema}, A. and {Bulik}, T. and {Bulten}, H.~J. and {Buonanno}, A. and {Buskulic}, D. and {Buy}, C. and {Byer}, R.~L. and {Cabero}, M. and {Cadonati}, L. and {Cagnoli}, G. and {Cahillane}, C. and {Calder{\'o}n Bustillo}, J. and {Callister}, T.~A. and {Calloni}, E. and {Camp}, J.~B. and {Campbell}, W.~A. and {Canepa}, M. and {Cannon}, K.~C. and {Cao}, H. and {Cao}, J. and {Carapella}, G. and {Carbognani}, F. and {Caride}, S. and {Carney}, M.~F. and {Carullo}, G. and {Casanueva Diaz}, J. and {Casentini}, C. and {Caudill}, S. and {Cavagli{\`a}}, M. and {Cavalier}, F. and {Cavalieri}, R. and {Cella}, G. and {Cerd{\'a}-Dur{\'a}n}, P. and {Cesarini}, E. and {Chaibi}, O. and {Chakravarti}, K. and {Chamberlin}, S.~J. and {Chan}, M. and {Chao}, S. and {Charlton}, P. and {Chase}, E.~A. and {Chassande-Mottin}, E. and {Chatterjee}, D. and {Chaturvedi}, M. and {Chatziioannou}, K. and {Cheeseboro}, B.~D. and {Chen}, H.~Y. and {Chen}, X. and {Chen}, Y. and {Cheng}, H. -P. and {Cheong}, C.~K. and {Chia}, H.~Y. and {Chiadini}, F. and {Chincarini}, A. and {Chiummo}, A. and {Cho}, G. and {Cho}, H.~S.},
        title = "{GW190425: Observation of a Compact Binary Coalescence with Total Mass {\ensuremath{\sim}} 3.4 M$_{{\ensuremath{\odot}}}$}",
      journal = {\apjl},
         year = 2020,
        month = mar,
       volume = {892},
       number = {1},
          eid = {L3},
        pages = {L3},
          doi = {10.3847/2041-8213/ab75f5},
archivePrefix = {arXiv},
       eprint = {2001.01761},
 primaryClass = {astro-ph.HE},
       adsurl = {https://ui.adsabs.harvard.edu/abs/2020ApJ...892L...3A}
}

@ARTICLE{2021A&A...645A...5S,
       author = {{Schneider}, F.~R.~N. and {Podsiadlowski}, Ph. and {M{\"u}ller}, B.},
        title = "{Pre-supernova evolution, compact-object masses, and explosion properties of stripped binary stars}",
      journal = {\aap},
         year = 2021,
        month = jan,
       volume = {645},
          eid = {A5},
        pages = {A5},
          doi = {10.1051/0004-6361/202039219},
archivePrefix = {arXiv},
       eprint = {2008.08599},
 primaryClass = {astro-ph.SR},
       adsurl = {https://ui.adsabs.harvard.edu/abs/2021A&A...645A...5S}
}
\bibliographystyle{aasjournal}

\end{document}